\documentclass[11pt]{article}

\usepackage[margin=1in]{geometry}
\usepackage{graphicx}
\usepackage{amsmath}
\usepackage{booktabs}
\usepackage{caption}
\usepackage[round,authoryear]{natbib}
\usepackage{xcolor}
\usepackage{hyperref}
\newcommand{\doi}[1]{\href{https://doi.org/#1}{doi:~#1}}

\graphicspath{{figures/}}

\title{Half-precision inference of AI weather prediction models: negligible loss of skill at a fraction of the cost}

\author{Stefan Siegert\footnote{Department of Mathematics and Statistics, University of Exeter, \texttt{s.siegert@exeter.ac.uk}. This work has not yet been peer-reviewed.}}
\date{6 November 2026}

\begin{document}

\maketitle

\begin{abstract}
AI weather prediction models are released with 32-bit (fp32)
weights and run in fp32 by default, although current GPUs compute much
faster in 16-bit formats. We run five publicly available models --
Aurora, Pangu-Weather, FourCastNet, SFNO and FastNet -- on a single
16\,GB GPU from 52 initial dates in 2024, in fp32 and with their networks
in half precision (fp16), and verify the forecasts against ERA5. The
physical state must stay in fp32: converting the whole model rounds it
in physical units, which in fp16 overflows and in bf16 discards
meaningful information. With the state, its normalisation and
denormalisation in fp32, running the network in fp16 changes the global
RMSE by at most 1.4\%, and by about 0.5\% or less for four of the five
models, equivalent to less than one hour of lead time up to day 8 and at
most two hours at day 10. The differences show no systematic pattern by
region, scale or for tropical and extratropical storms, apart from a
small tropical penalty in FourCastNet, and fp16 leaves the CRPS, spread
and calibration of Aurora's ensemble almost unchanged. No value
overflowed fp16's numeric range ($\pm65{,}504$); the largest stay 38 to 544 times below it. A
10-day forecast in fp16 is 1.3--3.5 times faster than in fp32 and uses
1.4--3.7 times less energy. Sensitivity experiments show that rules of
thumb from other applications of reduced precision carry over only
partly: the first and last layers dominate for some models but not for
others, and no type of layer is sensitive on its own. For running AI
weather models on a single GPU, fp16 with the state in fp32 is
practically lossless at a fraction of the cost. bf16, often preferred for
its wider range, is as fast but less accurate than fp16 for every model,
and simple 8-bit quantisation costs far more skill.
\end{abstract}

\section{Introduction}
\label{sec:introduction}

\subsection{Motivation}
\label{sec:intro-motivation}

 Within a few
years, weather forecasting models trained on reanalysis data have become
competitive with physics-based numerical weather prediction (NWP). Models
such as GraphCast \citep{lam2023graphcast}, Pangu-Weather
\citep{bi2023pangu} and Aurora \citep{bodnar2024aurora} match or exceed
the skill of ECMWF's high-resolution forecast on many standard
verification scores, and ECMWF now runs its own AI model, AIFS,
operationally \citep{lang2024aifs, ecmwf2026aifscard}. Independent assessments
against ECMWF's operational forecasts confirm that several of these models
match or exceed it on many headline scores, while differences remain, for
example in the intensity of extremes and the smoothness of the forecast
fields \citep{rasp2024weatherbench2, benbouallegue2024rise}.

Once trained, these models produce a 10-day forecast in seconds to
minutes on a single GPU or TPU, orders of magnitude faster than a
physics-based model on a supercomputer \citep{pathak2022fourcastnet,
bi2023pangu, lam2023graphcast}. Part of that gap reflects
the hardware rather than the method: ported to GPUs, physics-based models
run substantially faster than on CPUs \citep{fuhrer2018near}, by factors
of 6.4 node for node for ICON-A \citep{giorgetta2022icon} and 5.5
socket for socket for the operational ICON of MeteoSwiss
\citep{lapillonne2026icon}. Much of the remaining gap comes from the AI
models' longer time steps and coarser resolution; NeuralGCM, for example,
attributes its savings of three to five orders of magnitude to running at
8--40 times coarser horizontal resolution than the IFS and global
cloud-resolving models \citep{kochkov2024neural}. Cost nevertheless remains a constraint in research.
Studies that run a model many times -- over many initial dates, as
ensembles, or in sensitivity experiments -- multiply the cost of a
single forecast, and are often carried out on a single research computer
rather than an HPC cluster.

 The
size of a model that fits in GPU memory is often the binding constraint.
A consumer GPU with 16\,GB, such as the NVIDIA GeForce
RTX~5080 used in this study, launched at half the price of the model
with 32\,GB \citep{nvidia2025rtx50}; GPUs with more memory are
data-centre products, and access to large amounts of compute is
increasingly concentrated in industry rather than in academia
\citep{besiroglu2024compute}. Reducing the cost of running a model
also reduces its energy use, and inference -- repeated every time a
trained model is used -- is the larger part of that energy: at Google,
about three fifths of the energy used for machine learning in
2019--2021 went to inference rather than training
\citep{patterson2022carbon}, and the energy per inference varies by
orders of magnitude between models and tasks \citep{luccioni2024power}.

One of the simplest ways to reduce this cost is to run the model at
lower numerical precision. AI weather models, like most neural networks,
are distributed with their parameters (``weights'') stored as 32-bit
floating-point numbers (fp32), which carry about seven significant
decimal digits, and by default they are also run in fp32. The two common
16-bit formats, fp16 (half precision) and bfloat16 (bf16), carry about
three and two to three significant digits; fp16 is the more precise of
the two but represents only values up to 65{,}504, whereas bf16 keeps the
range of fp32. Storing a model in 16 bits halves the memory of its
weights, and recent GPUs, including consumer models, perform 16-bit
arithmetic considerably faster than fp32 arithmetic. Converting a trained
model to lower precision is commonly called \emph{quantisation}; the
term covers both reduced-precision floating-point formats and integer
formats, which need additional scaling. In this paper it refers to
running a model whose released weights are in fp32 with its network in
fp16 (or bf16). The formats, and the different ways a model can be run in
reduced precision, are described in \S\ref{sec:quantisation}.

Reduced precision is already common in AI weather models, but usually
as an implementation choice rather than as an object of study. Aurora
was trained with most computations in bf16, and for its air-pollution
extension its developers keep ``all computations before and after the
backbone'' in fp32 while the backbone runs in bf16, ``to better handle
low-magnitude values'' \citep{bodnar2024aurora}. The released code of
Aurora~1.5 runs most operations in a 16-bit format by default while
keeping the weights in fp32 -- a configuration that, at full
resolution, does not fit in the 16\,GB of GPU memory available here.
GraphCast's official code runs the network in bf16
\citep{lam2023graphcast, deepmind_weathernext_casting}; ECMWF's
inference software for AIFS offers 16-bit modes, noting that ``most
models work well with half precision'' \citep{anemoi_inference_precision};
and FourCastNet was trained with reduced precision at scale
\citep{kurth2023fourcastnet}. None of these sources reports how much
forecast skill reduced precision costs.

This paper asks what happens to the forecasts when an AI weather
model released with fp32 weights is run in 16 bits. Specifically: (i)
how much faster, smaller and more energy-efficient does the model
become; (ii) how much forecast skill does it lose, and does the loss
differ between models, variables, lead times, regions and scales; (iii)
is fp16's limited numeric range a practical risk; and (iv) where in the network
does the rounding enter the forecast, and do rules of thumb from other
applications of reduced precision -- such as keeping the first and last
layers or the normalisation layers in higher precision -- carry over to
weather models?

\subsection{Related work}
\label{sec:intro-related-work}

Physics-based weather and climate models have used reduced precision
for longer, and a common finding is that precision can be reduced
substantially once it is known \emph{where} in the model it is
needed. Large parts of a spectral dynamical core can run with
only a few significant bits \citep{dueben2014benchmark}. Single precision
in ECMWF's Integrated Forecasting System (IFS) reduced runtime by about
40\% without degrading forecast skill \citep{vana2017single} and has
been operational since IFS cycle 47R2, with the savings used to raise
the ensemble's vertical resolution from 91 to 137 levels
\citep{lang2021more}. Reduced precision has also been shown to be usable
in data assimilation \citep{hatfield2018improving}, to be tolerable for
small scales but not for large ones \citep{chantry2019scale}, to allow
hardware designed for deep learning to run parts of a weather model
\citep{hatfield2019accelerating}, and to allow a fluid model to run
entirely in 16 bits with suitable rounding and scaling
\citep{klower2022fluid, paxton2022climate}.

Machine-learning research shows that bf16 works well for training and
that sensitivity to reduced precision concentrates in a few parts of a
network. Training with a mix of 16-bit and 32-bit
arithmetic is well established \citep{micikevicius2018mixed}, and bf16
training matches the accuracy of fp32 training without changes to the
training procedure \citep{kalamkar2019bfloat16}. Whether a trained model
keeps its accuracy when it is afterwards run entirely in 16 bits is a
separate question: even a model trained with bf16 arithmetic, such as
Aurora, is released with fp32 weights, because standard
mixed-precision training keeps an fp32 master copy of the weights and
carries out the sums in normalisation and softmax layers in fp32
\citep{micikevicius2018mixed}. Research on quantisation shows that sensitivity to
reduced precision is unevenly distributed across a network. Image
classification networks are conventionally quantised with their first
and last layers kept at higher precision, ``as these have been reported
to significantly impact accuracy'' \citep{choi2018pact}. In large language
models, a small fraction of the internal values accounts for much of the
quantisation error, and keeping that fraction at higher precision recovers
most of the accuracy \citep{dettmers2022llmint8}. Weather models differ from these applications in one
respect that matters here: they are run autoregressively, each forecast
step taking the previous step's output as input, so that errors
introduced by reduced precision can grow over a forecast.

 The
closest work is a concurrent study of integer quantisation of two AI
weather models, DLWP and FourCastNet \citep{bhattacharya2026ptq}. It
converts the weights to 8-, 4- and 2-bit integers, with and without
quantising the intermediate results and with methods developed for
language models, evaluates 60-hour forecasts from four initial dates,
and finds that the number of bits used for the weights dominates the
loss of skill. Following common practice, it keeps the input and output
layers at full precision rather than testing their sensitivity, and
because the quantisation is simulated it does not measure speed or
memory gains. The present study differs in using formats that
current GPUs execute natively (fp16 and bf16), in measuring the speed,
memory and energy gains alongside the change in skill, and in covering
five models, 52 initial dates and lead times up to 10 days
(\S\ref{sec:intro-contributions}). It also tests where in each network
the rounding enters the forecast, including the sensitivity of the input
and output layers.

\subsection{Approach and contributions}
\label{sec:intro-contributions}

We run five publicly available AI weather models -- Aurora (in a
deterministic and an ensemble version), Pangu-Weather, FourCastNet,
SFNO and FastNet -- on a single 16\,GB GPU, from 52 initial dates in
2024, in fp32 and with their networks in fp16, and compare the forecasts
with each other and with ERA5. Our contributions are:
\begin{itemize}
\item We show that the physical state of a model must stay in fp32: the
simplest conversion of a whole model rounds it in physical units, which
in bf16 discards meaningful information and in fp16 overflows. Keeping
the state, its normalisation and denormalisation in fp32 adds little
memory or computation.
\item With the state in fp32, running the network in fp16 changes the
global RMSE by at most 1.4\% and for four of the five models by about
0.5\% or less, equivalent to losing less than an hour of lead time up to
day 8 and at most two hours at day 10, while making a forecast 1.3--3.5
times faster and 1.4--3.7 times cheaper in energy.
\item We examine where the fp16 and fp32 forecasts differ -- by region,
latitude band, spatial scale, in tropical and extratropical storms and in
an ensemble -- and find no systematic degradation beyond a small
tropical penalty in FourCastNet.
\item We show that fp16's limited numeric range is not a practical risk for
these models: no value of network internal activations or network outputs
overflowed, and the largest values stay 38 to 544 times below the limit.
\item We test where fp16 rounding enters the forecast and find that
common rules of thumb carry over only partly: the first and last layers
dominate for some models but not others, and no type of layer is
sensitive on its own.
\item We compare fp16 with bf16, the format often recommended because it
needs no attention to range: with the state in fp32, bf16 is as fast as
fp16 but less accurate for every model, by up to two orders of magnitude.
Simple 8-bit quantisation of the weights, or of weights and activations,
costs one to several orders of magnitude more skill than fp16.
\end{itemize}

\subsection{Outline}
\label{sec:intro-outline}

Section~\ref{sec:models-and-data} describes the models, their
skill and cost in fp32, the reduced-precision formats and the
verification data; Section~\ref{sec:methods} the experimental design and
diagnostics. Section~\ref{sec:results} presents the results on
quantisation: the need to keep the state in fp32, the comparison of fp16
with fp32 and where the two differ, ensembles, fp16's numeric range, where fp16
rounding enters the forecast, and the comparisons with bf16 and with
8-bit formats. Section~\ref{sec:discussion}
discusses the results and their limitations, and
Section~\ref{sec:conclusion} concludes.

\section{Models and Data}
\label{sec:models-and-data}

\subsection{Models}
\label{sec:models}

We study five AI weather models from three architecture families
(Table~\ref{tab:models}). They were chosen
from the models whose trained weights are publicly available and that
can be run in PyTorch on a single GPU with 16\,GB of memory, to cover
the main architecture families in current use: transformers (Aurora, Pangu-Weather), models that process
the fields in spectral space (FourCastNet, SFNO), and a graph neural
network (FastNet). All are
released with fp32 weights and run in fp32 by default; the fp32 forecast
is therefore the reference against which every reduced-precision
forecast is compared. All except FastNet run on the regular
0.25$^\circ$ latitude--longitude grid of ERA5, and all except
Pangu-Weather advance the forecast in 6-hour steps.

\begin{table}[htbp]
\centering
\footnotesize
\caption{Models studied. Variables: number of surface variables plus
number of atmospheric variables $\times$ number of pressure levels
(FourCastNet: 18 atmospheric fields on selected levels, see text; Aurora:
inputs, with seven further surface variables output only).
Runtime: time per forecast step in fp32 on our 16\,GB GPU
(Table~\ref{tab:benchmark}).}
\label{tab:models}
\setlength{\tabcolsep}{4pt}
\begin{tabular}{lllllrrl}
\toprule
Model & Version & Architecture & Grid & Step & Variables & Params & Runtime\\
\midrule
\shortstack[l]{FastNet\\\citep{dunstan2025fastnet}} & v1.1 & graph network & O96 ($\sim$1$^\circ$) & 6h & $7 + 5{\times}13$ & 62.2M & 0.29\,s \\
\shortstack[l]{FourCastNet\\\citep{pathak2022fourcastnet}} & v1 & Fourier operator & 0.25$^\circ$ & 6h & $8 + 18$ & 75.3M & 0.12\,s \\
\shortstack[l]{Pangu-Weather\\\citep{bi2023pangu}} & 24h model & 3D transformer & 0.25$^\circ$ & 24h & $4 + 5{\times}13$ & 276.7M & 0.79\,s \\
\shortstack[l]{SFNO\\\citep{bonev2023sfno}} & v0.1.0 & spherical Fourier op. & 0.25$^\circ$ & 6h & $8 + 5{\times}13$ & 289.4M & 0.53\,s \\
\shortstack[l]{Aurora\\\citep{bodnar2024aurora}} & 1.5 det./ens. & 3D Swin transformer & 0.25$^\circ$ & 6h & $18 + 5{\times}13$ & 1.26B & 5.4\,s$^\dagger$ \\
\bottomrule
\end{tabular}\\[2pt]
{\scriptsize $^\dagger$~With the backbone
weights streamed through GPU memory, the only way Aurora fits in 16\,GB
in fp32 (\S\ref{sec:quantisation}, \S\ref{sec:fp32-cost}).}
\end{table}

Aurora \citep{bodnar2024aurora} is a 1.26-billion-parameter model
pre-trained on a large and diverse set of weather and climate data and
then fine-tuned for forecasting. An encoder maps the input fields -- two
consecutive analysis times, 18 surface variables and five atmospheric
variables on 13 pressure levels -- to a compact internal representation;
a backbone of 3D Swin-transformer layers, which compute interactions
within local windows that shift from layer to layer, advances this
representation by one 6-hour step; and a decoder maps it back to the
physical variables. We use version~1.5 \citep{aurora_docs} in two
checkpoints from the same release: a deterministic one and an ensemble
one. The two have almost the same architecture and training data, but
the ensemble checkpoint adds random noise inside the model, so that
different random seeds give different ensemble members, and was fine-tuned
with the continuous ranked probability score (CRPS) as its loss instead
of the mean absolute error (MAE) \citep{weyn2026aurora15}.

Pangu-Weather \citep{bi2023pangu} is a 3D transformer with a
positional encoding adapted to the Earth's geometry, operating on four
surface variables and five atmospheric variables on 13 pressure levels.
Separate models were trained for 1-, 3-, 6- and 24-hour steps; we use
the 24-hour model, so its forecasts advance in 24-hour steps. The
official release provides the model only as exported ONNX computation
graphs, not as PyTorch code that can be converted to 16 bits component by
component, so we use a third-party PyTorch reimplementation of the
24-hour model with the official weights \citep{pangu_pytorch}.

FourCastNet \citep{pathak2022fourcastnet} (version~1) has 26 variables in the released
checkpoint (the original paper describes 20):
8 surface or single-level fields (10\,m and 100\,m wind, 2\,m temperature,
surface and mean sea-level pressure, total column water vapour) and 18
atmospheric fields, with each variable on its own subset of levels:
geopotential at 1000, 850, 500, 250 and 50\,hPa; the wind components at
1000, 850, 500 and 250\,hPa; temperature at 850, 500 and 250\,hPa; and
relative humidity at 850 and 500\,hPa. It
mixes information between grid points with an Adaptive Fourier Neural
Operator \citep{guibas2022afno}, which multiplies the Fourier transform
of the internal fields by learned complex-valued filters. The Fourier
transform treats the latitude--longitude grid as a flat plane rather
than a sphere. Although the filters are complex-valued, the
implementation stores their real and imaginary parts as separate
real-valued weights and carries out the complex multiplication with
real arithmetic, so the model can be converted to 16 bits, with one
small change to its code.

FastNet \citep{dunstan2025fastnet, daub2025fastnet}, developed by the
Met Office and the Alan Turing Institute (version~1.1), is a graph
neural network of the GraphCast family \citep{lam2023graphcast}. Its
encoder passes information from the grid to an icosahedral mesh of
10{,}242 nodes in a single message-passing step; its processor performs
16 rounds of message passing on that mesh, each with its own weights;
and its decoder passes the result back to the grid, adding it to the
input state. It forecasts seven surface variables and five atmospheric
variables on 13 pressure levels, with 12 further static and time-of-day
fields as input. Unlike the other models, it runs on ECMWF's O96 reduced
Gaussian grid (40{,}320 points, roughly 1$^\circ$ resolution). We build
its initial conditions from the same 0.25$^\circ$ ERA5 data as for the
other models, by linear interpolation onto the O96 points, and
interpolate its forecasts back to the 0.25$^\circ$ grid before
verification (\S\ref{sec:verification-data}), so that all models are
scored on the same grid against the same truth.

SFNO \citep{bonev2023sfno} (the \texttt{sfno\_73ch\_small}
checkpoint, version~0.1.0) replaces FourCastNet's flat Fourier transform
with a transform in spherical harmonics, which respects the geometry of
the sphere. It forecasts eight surface variables and five atmospheric
variables on 13 pressure levels. Its spectral filters, 97.8\% of its
parameters, are complex-valued; in 16 bits we apply them as real and
imaginary parts, as FourCastNet does.

We also considered two further models. ECMWF's AIFS \citep{lang2024aifs}
(\texttt{aifs-single-2.0}) runs on our GPU, but it needs six ocean-wave
variables as input that ERA5 does not archive, so it cannot be
initialised like the other models. GraphCast \citep{lam2023graphcast} is
implemented in JAX, and no PyTorch version loads its official weights.

Like most deep-learning models, AI weather models can be divided into
components. In a forecast step, the physical fields of the state are
typically normalised, passed through an encoder, a number of backbone
layers or blocks, and a decoder, and the decoder's output is
denormalised into the physical fields of the next state. We use this
division in the experiments that locate where fp16 rounding enters the
forecast (\S\ref{sec:methods-tier2}). The components differ widely in
their numbers of parameters (Table~\ref{tab:components}).

\begin{table}[htbp]
\centering
\small
\caption{Components per model, in forward-pass order, with parameter counts and share of the model total. Repeated identical blocks are grouped. ``Other'': parameters outside every component, which stay fp32 in the one-at-a-time experiments (\S\ref{sec:methods-tier2}) (FourCastNet: its learned positional embedding; Aurora: backbone time/noise MLPs and context-downsampling layers; SFNO: the $1\times1$ convolution on its residual path). SFNO's complex filter weights are counted once each.}
\label{tab:components}
\begin{tabular}{llrr}
\toprule
Model & Component & Parameters & Share \\
\midrule
Aurora & \texttt{encoder} & 7{,}665{,}152 & 0.608\% \\
 & \texttt{backbone\_stage\_0} & 27{,}307{,}008 & 2.17\% \\
 & \texttt{backbone\_stage\_1} & 155{,}330{,}560 & 12.3\% \\
 & \texttt{backbone\_stage\_2} & 436{,}420{,}608 & 34.6\% \\
 & \texttt{backbone\_stage\_3} & 445{,}859{,}840 & 35.4\% \\
 & \texttt{backbone\_stage\_4} & 149{,}294{,}080 & 11.8\% \\
 & \texttt{backbone\_stage\_5} & 25{,}205{,}760 & 2\% \\
 & \texttt{decoder} & 9{,}953{,}776 & 0.79\% \\
 & other & 3{,}147{,}776 & 0.25\% \\
 & \textit{total} & 1{,}260{,}184{,}560 & \\
\addlinespace
Pangu-Weather & \texttt{patch\_embedding} & 58{,}752 & 0.0212\% \\
 & \texttt{stage\_0} & 31{,}744{,}896 & 11.5\% \\
 & \texttt{downsample} & 296{,}448 & 0.107\% \\
 & \texttt{stage\_1..2 (2$\times$)} & 106{,}198{,}272 & 38.4\% each \\
 & \texttt{upsample} & 332{,}160 & 0.12\% \\
 & \texttt{stage\_3} & 31{,}744{,}896 & 11.5\% \\
 & \texttt{patch\_recovery} & 86{,}240 & 0.0312\% \\
 & \textit{total} & 276{,}659{,}936 & \\
\addlinespace
FourCastNet & \texttt{patch\_embed} & 1{,}278{,}720 & 1.7\% \\
 & \texttt{block\_0..11 (12$\times$)} & 5{,}023{,}488 & 6.67\% each \\
 & \texttt{head} & 1{,}277{,}952 & 1.7\% \\
 & other & 12{,}441{,}600 & 16.5\% \\
 & \textit{total} & 75{,}280{,}128 & \\
\addlinespace
SFNO & \texttt{encoder} & 177{,}408 & 0.0613\% \\
 & \texttt{block\_0..7 (8$\times$)} & 36{,}129{,}408 & 12.5\% each \\
 & \texttt{decoder} & 175{,}872 & 0.0608\% \\
 & other & 5{,}621 & 0.0019\% \\
 & \textit{total} & 289{,}394{,}165 & \\
\addlinespace
FastNet & \texttt{mesh\_edge\_embedder} & 594{,}432 & 0.955\% \\
 & \texttt{encoder} & 1{,}843{,}200 & 2.96\% \\
 & \texttt{processor\_0..15 (16$\times$)} & 3{,}545{,}088 & 5.7\% each \\
 & \texttt{decoder} & 3{,}077{,}448 & 4.94\% \\
 & \textit{total} & 62{,}236{,}488 & \\
\bottomrule
\end{tabular}
\end{table}

\begin{figure}[htbp]
\centering
\includegraphics[width=\textwidth]{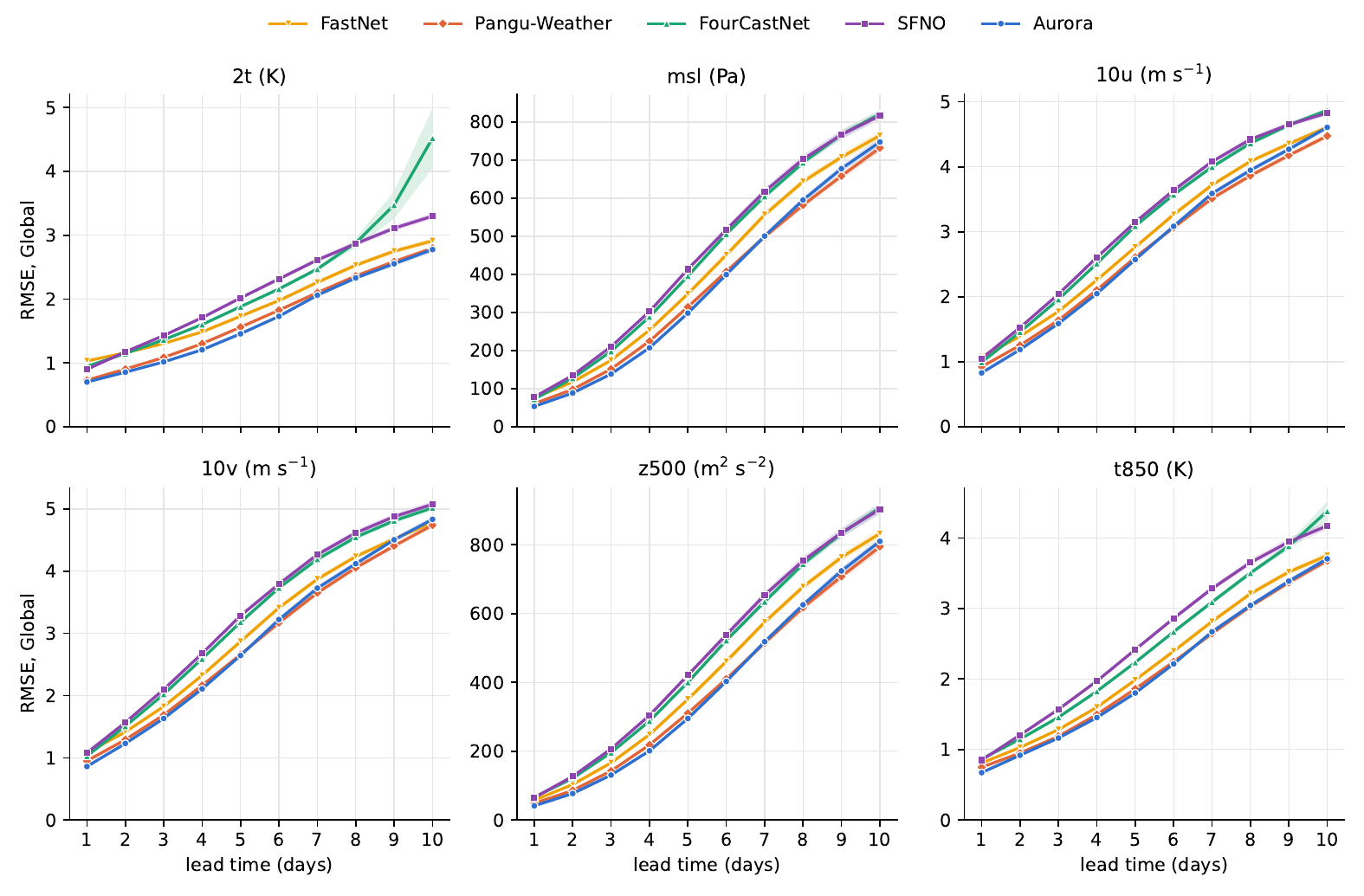}
\caption{fp32 forecast skill of the five models (Aurora: deterministic checkpoint): mean global RMSE against ERA5 at daily lead times, 52 initialisations in 2024, all models verified on the regular 0.25$^\circ$ grid (FastNet's forecasts linearly interpolated from its O96 grid). Shading: $\pm$1 standard error across dates. UK/Europe: Figure~\ref{fig:model-skill-uk} in
Appendix~\ref{sec:appendix-figures}.}
\label{fig:model-skill}
\end{figure}

In fp32, Aurora's deterministic checkpoint is the most accurate model
at 1 day, Pangu-Weather is level with it by 10 days, and FourCastNet and
SFNO are the least accurate from day 5 on (Figure~\ref{fig:model-skill};
all models verified on the same 52 initialisations, ERA5 truth and
0.25$^\circ$ grid, \S\ref{sec:verification-data}). At 1 day, Aurora has
the lowest global RMSE for every variable, with Pangu-Weather second; at
10 days, Pangu-Weather is level with or marginally better than Aurora for
every variable. FourCastNet and SFNO have the highest RMSE at days 5 and
10 for every variable; FourCastNet's 2t error grows most by day 10, where
a few of its fp32 forecasts blow up (\S\ref{sec:results-fp16}). FastNet
is among the least skilful models at day 1 (last for 2t and 10u) but
second or third by day 10. This is plausibly a resolution effect: its native grid
is about 1$^\circ$, and interpolation to 0.25$^\circ$ cannot restore
small-scale detail, which matters most at short lead times. Aurora's
ensemble checkpoint is verified as an ensemble in
\S\ref{sec:results-ensemble}. Figure~\ref{fig:model-skill-uk}
(Appendix~\ref{sec:appendix-figures}) shows the same comparison for the
UK/Europe region.

\label{sec:fp32-cost}
In fp32, a forecast step takes 0.12\,s (FourCastNet), 0.29\,s
(FastNet), 0.53\,s (SFNO), 0.79\,s (Pangu-Weather, a 24-hour step) and
5.4\,s (Aurora) on our GPU (Table~\ref{tab:benchmark},
\S\ref{sec:methods-further}), so that a 10-day forecast takes between 5\,s
and 3.6\,minutes and uses between 0.4 and 18\,Wh of energy. Peak memory
ranges from 1.2\,GB (FourCastNet) to 15.4\,GB (SFNO), which leaves SFNO
little headroom on a 16\,GB GPU. Aurora does not fit in 16\,GB in fp32 at
all: its fp32 runs move the backbone's weights to the GPU one block at a
time (\S\ref{sec:quantisation}), which keeps its peak memory at 14.3\,GB
but makes it the slowest model by far. These costs are the reference for
the reduced-precision runs (\S\ref{sec:results-fp16}).

\subsection{Quantisation}
\label{sec:quantisation}

A floating-point number is stored as a sign, an exponent and a
significand (mantissa). fp32 uses 8 exponent and 23 mantissa bits, so
that rounding a value to fp32 changes it by at most about
$6\times10^{-8}$ of its magnitude. The two common 16-bit formats divide
their bits differently. fp16 uses 5 exponent and 10 mantissa bits:
rounding changes a value by at most $2^{-11}\approx0.05\%$, but values
above 65{,}504 cannot be represented and become infinite, and values below
$6.1\times10^{-5}$ lose precision. bf16 keeps fp32's 8 exponent bits, and
hence its range (about $10^{-38}$ to $3\times10^{38}$), but only 7
mantissa bits, so that rounding changes a value by up to
$2^{-8}\approx0.4\%$, eight times more than fp16. bf16 can therefore
replace fp32 without regard to the size of the values
\citep{kalamkar2019bfloat16}; fp16 is more precise, but only if the values
stay within its range.

The simplest way to run a model in 16 bits is to convert the whole
model with one command (in PyTorch, \texttt{model.to(torch.bfloat16)} or
\texttt{model.to(torch.float16)}). This converts the weights, but a layer
can only multiply inputs that have the same precision as its weights, so
the inputs must be converted too. In models that normalise their inputs
themselves, this happens before the normalisation: Aurora's code
converts the input fields to the precision of its weights and then
normalises them, and Pangu-Weather normalises inside its first layer,
which therefore receives 16-bit inputs. The physical fields themselves
are then rounded to 16 bits at every step, which in fp16 overflows and in
bf16 costs substantial skill (\S\ref{sec:results-state}). In all our
16-bit runs, the state, its normalisation and denormalisation, including
the normalisation statistics, therefore stay in fp32, and only the
network is converted.

In the \emph{fp16} and \emph{bf16} runs, every weight and every
computation of the network is in that format, while the normalisation of
the inputs, the denormalisation of the outputs and the state carried from
one forecast step to the next stay in fp32. The normalised fields are rounded to 16
bits when they enter the network, and its output is converted back to
fp32 before it is denormalised. For Aurora, the input fields are kept and
normalised in fp32 and rounded on entry to the encoder; for
Pangu-Weather, the normalisation is moved out of the first layer; FastNet
carries a normalised state, which is kept in fp32 between steps;
FourCastNet's code already normalises and denormalises outside the
network in fp32; and for SFNO, the solar zenith angle and static fields
are added to the state in fp32 before it enters the network. We verified that the state never reaches the
network in 16 bits, and, for Pangu-Weather, that the modified code with
fp32 weights reproduces the original model bit for bit. Every weight is
rounded once -- in bf16, Aurora's weights change by 0.14\% on average --
and every intermediate result inside the network is stored in 16 bits.
Because current GPUs sum the products in a 16-bit matrix multiplication
in fp32, the main source of error within the network is the rounding of
these stored values rather than the arithmetic within an operation. A
few operations are computed in fp32 on 16-bit inputs: FourCastNet's
Fourier transforms and SFNO's spherical harmonic transforms. FourCastNet's network is converted component by
component (Table~\ref{tab:components}), which leaves its positional
embedding, part of no component, in fp32. Further numerical details are
given in the code's \texttt{nwpprec/README.md} \citep{siegert2026code}.

The fp32 reference is each model as PyTorch runs it by default. This
includes one deviation from strict fp32 that we kept deliberately:
PyTorch computes convolutions, which occur in the input and output
layers of Aurora, Pangu-Weather and FourCastNet, in the reduced-precision
format TF32 by default. Aurora
does not fit in 16\,GB of GPU memory in fp32; its fp32 runs therefore
move the backbone weights to the GPU one block at a time and compute
the decoder in spatial portions. This makes them slower
(\S\ref{sec:fp32-cost}) but, as we verified on test data, gives
bit-identical results.

The approach used here is a \emph{persistent} conversion: the model is
converted once and the fp32 weights are discarded. The most common
alternative, automatic mixed precision \citep{micikevicius2018mixed}, is the
default inference mode of some models, such as Aurora~1.5. It keeps the
weights in fp32 and converts values to 16 bits only for individual operations.
This protects accuracy but saves little memory: in this mode, Aurora~1.5 runs
out of memory on our 16\,GB GPU in the first forecast step, whereas the
persistently converted model needs at most 11.6\,GB
(Table~\ref{tab:benchmark}). The persistent conversion is the harsher of the
two for accuracy, so the costs reported here are those of the configuration
that fits on such a GPU.

A larger literature concerns quantisation to integers (int8, int4),
in which values are represented as integers multiplied by a scale shared
by a tensor, a channel or a group of values
\citep{jacob2018quantization, nagel2021white}. The scales are chosen from calibration data
after training (post-training quantisation) or learned during training
(quantisation-aware training), and either the weights alone or also the
intermediate results are quantised. \citet{bhattacharya2026ptq} applied
such methods, including ones developed for language models
\citep{xiao2022smoothquant, lin2023awq}, to two AI weather models.
8-bit floating-point formats (fp8) also need scaling
\citep{micikevicius2022fp8}. We did not pursue these formats: they need
calibration and low-precision implementations of every operation, and in
tests with Aurora they were either slower than bf16 or did not fit in
16\,GB. fp16 and bf16 are the
reduced-precision formats that most users of these models can apply
directly: they need no calibration, and fast fp16 arithmetic is
available on NVIDIA GPUs from 2017 on, bf16 arithmetic from 2020 on
\citep{nvidia2017volta, nvidia2020ampere}, and both on current consumer and
data-centre models. On CPUs, by contrast, fast fp16 arithmetic is
not usually available, so the speedups reported here should not be
expected to carry over to them.

\subsection{Model verification}
\label{sec:verification-data}

All forecasts are initialised from, and verified against, the ERA5
reanalysis \citep{hersbach2020era5} on its regular 0.25$^\circ$
latitude--longitude grid ($721\times1440$ points), the reference
commonly used to evaluate AI weather models
\citep{rasp2020weatherbench}. Models that take two input times (Aurora)
receive the analyses at the initial time and six hours earlier.
The grids of Aurora and FourCastNet omit the southernmost row of the ERA5
grid, which is therefore not scored for these models. Aurora~1.5 was fine-tuned last on
operational IFS analyses rather than ERA5 \citep{aurora_docs}; starting
it from ERA5 may lower its fp32 skill slightly, but affects its fp32
and fp16 forecasts alike.

Initial dates are drawn from 2024, a year after the training periods of
the models for which these are documented (\S\ref{sec:methods-tier1}
describes the sampling). 2024 lies after the training periods of
Pangu-Weather (1979--2017; \citealp{bi2023pangu}), FourCastNet
(1979--2015; \citealp{pathak2022fourcastnet}) and FastNet (1980--2021;
\citealp{daub2025fastnet}), and was chosen over 2023 because Aurora's
training data include 2023: Aurora~1.5 was trained on ERA5 from
1981--2023 and fine-tuned on operational analyses from 2018--2023, with
2024 held out as its test year \citep{weyn2026aurora15}. The SFNO
checkpoint's documentation gives no training period; the SFNO paper,
which it cites, trains its 73-variable models on 1979--2015 with 2016 and
2017 for validation \citep{bonev2023sfno}, the same split into training,
validation and out-of-sample data that the checkpoint's stored
configuration records, and the developer's SFNO inference service lists ERA5
from 1979--2017. Either way, 2024 lies outside it.

FastNet runs on the O96 grid of about 1$^\circ$ resolution
(\S\ref{sec:models}). Its initial conditions are obtained by linear
interpolation of the 0.25$^\circ$ ERA5 fields onto the O96 points, and
its forecasts are interpolated back to the 0.25$^\circ$ grid by linear
interpolation on the Delaunay triangulation of the O96 points, so that
all models are scored on the same grid against the same truth. The grid
points outside the triangulation, about 1.6\% (a gap at the 0$^\circ$
meridian and the rows at the poles), have no interpolated value and are not
scored for FastNet. Every
FastNet score in this paper goes through this step.

We verify 2\,m temperature (2t), mean sea-level pressure (msl),
10\,m wind speed (ws10, computed from the two wind components),
geopotential at 500\,hPa (z500) and temperature at 850\,hPa (t850), at
daily lead times from 1 to 10 days; Figure~\ref{fig:model-skill} shows
the two wind components instead of the wind speed. Since the lead times
are multiples of 24 hours, all models, including Pangu-Weather with its
24-hour steps, are verified at the same times. Each forecast is scored
globally, over three regions and at five locations
(\S\ref{sec:methods-regions}).

The main score is the root-mean-square error (RMSE) of each forecast
against ERA5, with each grid point weighted by the cosine of its
latitude to account for the convergence of the meridians
\citep{rasp2020weatherbench}. We also record the bias and the mean
absolute error (MAE), the deterministic counterpart of the CRPS. Scores
are computed per forecast and averaged over initial dates. The ensemble
forecasts (\S\ref{sec:results-ensemble}) are additionally scored by the
area-weighted mean of the fair continuous ranked probability score
(CRPS), which does not penalise ensembles for having few members
\citep{ferro2014fair}; by the ensemble spread, the square root of the
area mean of the ensemble variance, compared with the RMSE of the
ensemble mean after correcting for the finite ensemble size
\citep{leutbecher2008ensemble}; and by the fraction of grid points at
which ERA5 lies outside the range of the ensemble, which for a
calibrated ensemble of $M$ members is $2/(M+1)$ \citep{siegert2011outliers}.

\section{Methods}
\label{sec:methods}

\subsection{Regions and locations}
\label{sec:methods-regions}

Every forecast is scored globally, for three regions of
similar size (Table~\ref{tab:regions}) and for five metropolitan areas,
each represented by the $3\times3$ grid points
($0.75^\circ\times0.75^\circ$) nearest to its centre. The regions cover the
mid-latitudes of two continents and a monsoon region; North America is the
WMO's standard verification area \citep{wmo2010gdpfs}. The
locations span oceanic, monsoon, Mediterranean, continental and
subtropical-highland climates in both hemispheres. The Bay Area box
includes points over the bay and the Pacific. In addition, error sums on
$10^\circ$ cells give latitude bands
(\S\ref{sec:methods-further}).

\begin{table}[tbp]
\centering
\small
\caption{Regions and locations used for verification. Locations: the
$3\times3$ grid points nearest to the given centre.}
\label{tab:regions}
\begin{tabular}{lll}
\toprule
Name & Area & Climate \\
\midrule
UK/Europe & 35--72$^\circ$N, 15$^\circ$W--40$^\circ$E & mid-latitude \\
North America & 25--60$^\circ$N, 145--50$^\circ$W & mid-latitude, continental \\
South Asia & 5--35$^\circ$N, 60--100$^\circ$E & tropical and subtropical monsoon \\
\midrule
London & 51.5$^\circ$N, 0.1$^\circ$W & temperate oceanic \\
Delhi & 28.6$^\circ$N, 77.2$^\circ$E & semi-arid, monsoon \\
Bay Area & 37.6$^\circ$N, 122.2$^\circ$W & Mediterranean, coastal \\
Beijing & 39.9$^\circ$N, 116.4$^\circ$E & continental monsoon \\
S\~{a}o Paulo & 23.55$^\circ$S, 46.65$^\circ$W & subtropical highland \\
\bottomrule
\end{tabular}
\end{table}

\subsection{Paired fp32 and fp16 forecasts}
\label{sec:methods-tier1}

For each model and initial date, we run the same 10-day forecast from
identical initial conditions with the network in fp32 and in fp16
(\S\ref{sec:quantisation}), and score both against ERA5
(\S\ref{sec:verification-data}) at daily lead times (days 1--10). Because
the two forecasts share the initial state and the weather situation,
their difference in RMSE isolates the effect of precision from the much
larger variation of forecast error between dates. We also record the RMS
difference between the fp16 and the fp32 forecast, their
\emph{departure}, which measures how much the precision changes the
forecast regardless of whether it becomes more or less accurate. The bf16
forecasts used for the secondary comparison (\S\ref{sec:results-bf16})
are built and scored in the same way.

The initial dates are drawn from 2024 by stratified random sampling:
one date from each of the 52 calendar weeks, with the initialisation
times 00, 06, 12 and 18\,UTC used 13 times each and the days of the week
used as evenly as possible. All models use the same 52
initialisations, so that differences between them cannot arise from
differences in the weather situations they were tested on. Initialisations about a week
apart give forecasts whose errors are close to independent between
dates. The sensitivity experiments and the ensemble
(\S\ref{sec:methods-tier2}, \S\ref{sec:methods-further}) use every fourth
of these dates (13).

For each model, variable, lead time and area, we report the change in
mean RMSE from fp32 to fp16 relative to the fp32 RMSE, with a 95\%
confidence interval and a two-sided paired $t$-test of the per-date RMSE
differences at the 5\% level. With 50 combinations of variable and lead
time per model and area, we do not correct for multiple testing, and
treat isolated significant results, especially at the locations, with
caution. To judge whether a change matters in practice, we also express
it as the lead time over which the fp32 RMSE grows by the same amount
(\S\ref{sec:results-fp16}).

To show what this design can detect, we compute for each combination
the minimum detectable effect \citep{bloom1995minimum,cohen1988statistical}:
the smallest change in mean RMSE that the paired $t$-test detects with
80\% probability, given $N=52$ dates and the observed standard deviation
$s$ of the per-date differences, $(t_{0.975,N-1}+t_{0.8,N-1})\,s/\sqrt{N}$
(Appendix~\ref{sec:appendix-mde}, Table~\ref{tab:power}). This is based on the observed variability, not
on the observed effect, so it describes the sensitivity of the design
rather than the probability that the reported effects are real.
Globally, the median detectable change is 0.03--0.33\% of the fp32 RMSE,
depending on the model; over the three regions, where scores vary more
from date to date, 0.08--1.1\%. Because the fp16 and fp32 forecasts are
so closely paired, the design resolves changes far smaller than any that
would matter in practice: a significant result here shows that fp16
changes the forecast, not that the change is important.

\begin{table}[htbp]
\centering
\small
\caption{Smallest change in mean RMSE, as \% of the fp32 RMSE, that the paired $t$-test detects with probability 0.8 at the 5\% level, given the $N$ dates and the observed date-to-date variability of the fp16$-$fp32 differences. Median over the 50 combinations of variable and lead time per area (global: with range).}
\label{tab:power}
\begin{tabular}{lrrrrr}
\toprule
Model & $N$ & Global & UK/Europe & N.\ America & S.\ Asia \\
\midrule
FastNet & 52 & 0.33 [0.01, 0.7] & 1.11 & 1.00 & 0.52 \\
Pangu-Weather & 52 & 0.05 [0.01, 0.2] & 0.16 & 0.13 & 0.14 \\
FourCastNet & 52 & 0.23 [0.02, 6.5] & 0.95 & 0.66 & 1.13 \\
SFNO & 52 & 0.03 [0.00, 0.1] & 0.11 & 0.08 & 0.09 \\
Aurora & 52 & 0.12 [0.01, 0.5] & 0.36 & 0.30 & 0.54 \\
\bottomrule
\end{tabular}
\end{table}

\subsection{Sensitivity experiments}
\label{sec:methods-tier2}

To find where in the network fp16 rounding enters the forecast, we
use a one-at-a-time (OAT) design on the components listed in
Table~\ref{tab:components}: one component at a time runs in fp16 and the
rest of the network in fp32, which measures the departure that the
component causes on its own. The inputs of the component are rounded to
fp16 when they enter it, and its outputs converted back to fp32 when they
leave it. This corresponds to the noising variant of activation patching, in which a
component is perturbed in an otherwise clean model
\citep{heimersheim2024activation}, a technique used to locate computations
in language models \citep{meng2022locating}. One-at-a-time designs cannot detect
interactions between components \citep{saltelli2010perfunctory}; we
check for them by comparing the components' effects with that of the
whole network (\S\ref{sec:results-sensitivity}), and we also run the
network with every component converted in this way.

Each forecast is compared with the fp32 forecast from the same
initial conditions rather than with ERA5: the departure, the
area-weighted RMS difference between the two, is computed globally for
each variable at every model step up to 24\,h and averaged over the 13
dates. A comparison with ERA5 would add the much larger forecast error
that both forecasts share. We express the departure $D_c$ with component
$c$ in fp16 relative to the departure $D_\mathrm{fp16}$ of the whole
network in fp16, as the share $D_c/D_\mathrm{fp16}$ of the departure that
the component causes on its own.

The one-at-a-time design asks where in the network the rounding
matters. Practice in other applications of reduced precision suggests
three other ways in which sensitivity could concentrate: in certain types
of operation, such as normalisation and softmax, whose sums mixed-precision
training carries out in fp32 \citep{micikevicius2018mixed}; in a few channels
with outsized values, which dominate the quantisation error of large
language models \citep{dettmers2022llmint8}; and towards the end of the
network, if rounding errors accumulate with depth. Whether these carry
over to weather models determines whether rules from those fields can
guide reduced-precision inference here, so we test each on the same 13
dates. (i) All layers of one type -- normalisation, attention or spectral
filters, MLPs (FastNet: normalisation and linear layers) -- are kept in
fp32 with their original weights inside the fp16 network, with fp16
activations passed between them. (ii) The largest magnitude of every
output channel of every module is recorded in the fp16 forecasts. (iii)
The relative difference between the fp16 and fp32 outputs of every module
is computed during the first model step, from a fixed sample of $2^{14}$
values of each output.
FastNet runs in PyTorch's
deterministic mode in all these experiments, so that its departures
contain no run-to-run variation.

\subsection{Further diagnostics}
\label{sec:methods-further}

Latitude bands and maps (\S\ref{sec:results-where})
are computed from cos-latitude-weighted sums of the squared error, the
error and the squared departure from the fp32 forecast, accumulated on
$10^\circ\times10^\circ$ cells for every forecast, variable and lead
time; RMSE, bias and departure can then be recomputed exactly for any
union of cells.

To see whether fp16 smooths or roughens the forecast fields, we
compute, in the same runs, the zonal power spectrum of each forecast
field and of ERA5 -- the squared magnitude of its Fourier transform along
each latitude circle -- averaged with cos-latitude weights over each
$10^\circ$ latitude band. We compare the spectra of each model up to the
smallest scale it resolves, set by its patch size (Aurora and Pangu-Weather),
the wavenumbers its spectral filters act on (FourCastNet), its spherical
harmonic truncation (SFNO) or its grid spacing (FastNet).

Aurora's ensemble checkpoint is run on every fourth of the 52
dates (13) with 8 members, whose noise is set by seeds 0--7 in both fp32
and fp16, so that member $m$ of the fp16 ensemble is the same stochastic
draw as member $m$ of the fp32 ensemble. Each ensemble is scored, for
2t, msl, z500 and t850, by the
fair CRPS \citep{ferro2014fair}, the RMSE of the ensemble mean, the spread
(square root of the area-mean ensemble variance) and the fraction of the
area where ERA5 lies outside the range of the members; the precision effect
is the RMS difference between same-seed fp16 and fp32 members.

To test whether fp16's numeric range suffices, the fp16 forecasts of
all 52 dates are repeated with forward hooks on every module of each
network (not only the innermost layers, so that sums formed inside a
block are seen at its output). For every fp16 tensor a module returns,
they record the largest finite magnitude, the number of infinite or
undefined values, and the number of nonzero values below
$6.1\times10^{-5}$, both per module and per model step. Values computed
and used within a module without being returned, such as attention
scores, are not seen directly, but an overflow there would make the
module's output infinite or undefined.

For a qualitative view of high-impact weather, we track four named
storms of 2024 whose initialisation is one of the 52 dates: the
extratropical storms Isha (initialised 20 January 00\,UTC;
\citealp{metoffice2024isha}) and Kathleen (5 April 18\,UTC;
\citealp{metoffice2024kathleen}), and the hurricanes Beryl (1 July 00\,UTC;
\citealp{beven2025beryl}) and Milton (7 October 00\,UTC;
\citealp{beven2025milton}). Each storm is followed in ERA5 and in each forecast
by the minimum of mean sea-level pressure within 700\,km (extratropical)
or 400\,km (tropical) of its previous position per 6 hours, starting from
its ERA5 position (for Isha, from 21 January 12\,UTC, when it has become a
separate low). We compare track position, central pressure and the
maximum 10\,m wind within 300\,km of the centre with ERA5, for fp32 and
fp16. ERA5 underestimates the
intensity of tropical cyclones, so intensities are compared with ERA5,
not with observed intensities.

The cost of each model and precision is measured on the GPU used
throughout, an NVIDIA GeForce RTX~5080 with 16\,GB of memory, with
nothing else running on it. Each model is stepped forward from one
initial date, restarting the 10-day forecast as often as needed, until at
least 60\,s and at least 8 steps have been timed; the first step of each
forecast, which also loads the inputs, is not timed. We report the mean
time per step, the time for a 10-day forecast, the peak GPU memory
allocated by PyTorch, and the energy of a 10-day forecast: the mean board
power of the GPU, sampled every 50\,ms within the timed steps, times
their duration.

To test whether 8-bit formats are a further option, we simulate
8-bit quantisation inside the fp16 network, with the state in fp32, on
the 13 sensitivity dates over 10 days. In the weight-only configurations,
every weight of a matrix product -- the linear and convolution layers and
FourCastNet's and SFNO's spectral filters (for SFNO, the real and
imaginary parts) -- is rounded to int8 (symmetric, 127 levels) or to fp8
(e4m3) with one scale per output channel, and the network then computes
in fp16. In the configurations with 8-bit inputs, the inputs of every
linear and convolution layer are also rounded, to int8 with one scale per
token or to fp8 with one scale per tensor. Biases, normalisation layers,
embeddings and Pangu-Weather's position-dependent attention bias, which
holds 91\% of its parameters, stay in fp16, as is usual. This is the
simplest form of 8-bit quantisation, round-to-nearest without
calibration or the treatment of outliers developed for language models
\citep{dettmers2022llmint8, xiao2022smoothquant, lin2023awq}; it
reproduces the accuracy of 8-bit inference but not its speed or memory,
which we do not measure.

\section{Results}
\label{sec:results}

\subsection{The physical state must stay in fp32}
\label{sec:results-state}

Before comparing fp16 and fp32 networks, we show why the
physical state must be kept out of the conversion: the simplest way to
run a model in 16 bits, converting the whole model, also rounds the
state, and this alone degrades or destroys the forecasts. Converted as a
whole to 16 bits, Aurora and Pangu-Weather hold
their state -- including the output of each step, which is the input of
the next -- in 16 bits in physical units, because they convert their
input fields before normalising them (\S\ref{sec:quantisation}).
FourCastNet and FastNet normalise outside the network and
are not affected; SFNO's whole-model conversion fails differently, by
discarding the imaginary parts of its filters.

\begin{table}[tbp]
\centering
\small
\caption{Rounding the physical state to 16 bits: for ERA5 initial states (13 dates in 2024), the largest magnitude, the share of values above fp16's largest value (65,504), and the RMS error of rounding each field to bf16 and fp16 in physical units (fp16: over the values that stay finite) and after normalisation with the field's global mean and standard deviation (error expressed in physical units).}
\label{tab:state-rounding}
\begin{tabular}{lrrrrrr}
\toprule
 & & & \multicolumn{2}{c}{physical units} & \multicolumn{2}{c}{normalised} \\
Field & max & \% $>$ 65{,}504 & bf16 & fp16 & bf16 & fp16 \\
\midrule
2t (K) & 323 & 0 & 0.565 & 0.0708 & 0.0255 & 0.00319 \\
t850 (K) & 314 & 0 & 0.567 & 0.0711 & 0.0202 & 0.00252 \\
msl (Pa) & 1.05e+05 & 100 & 148 & overflow & 1.88 & 0.234 \\
sp (Pa) & 1.05e+05 & 97 & 148 & overflow & 11.3 & 1.41 \\
z500 (m$^2$\,s$^{-2}$) & 5.89e+04 & 0 & 73.9 & 9.24 & 4.59 & 0.573 \\
z400 (m$^2$\,s$^{-2}$) & 7.56e+04 & 81 & 143 & overflow & 5.64 & 0.705 \\
z300 (m$^2$\,s$^{-2}$) & 9.67e+04 & 100 & 148 & overflow & 6.92 & 0.864 \\
z50 (m$^2$\,s$^{-2}$) & 2.08e+05 & 100 & 295 & overflow & 6.22 & 0.777 \\
\bottomrule
\end{tabular}
\end{table}

Table~\ref{tab:state-rounding} shows what 16 bits do to the
physical state. In fp16, every value of mean sea-level pressure in Pa, 97\%
of surface pressure and all geopotential above 400\,hPa (81\% at 400\,hPa)
exceed 65{,}504 and become infinite. bf16 represents them, but coarsely:
rounding an ERA5 initial state to bf16 changes 2t and t850 by 0.57\,K, msl
by 148\,Pa and z500 by 74\,m$^2$\,s$^{-2}$ (RMS), comparable to the models'
fp32 errors at day 1 (0.70--1.03\,K for 2t, 54--78\,Pa for msl,
41--66\,m$^2$\,s$^{-2}$ for z500; \S\ref{sec:models}), and the rounding
recurs at every step. The same fields rounded after normalisation, whose
values are of order one, change by 16 to 80 times less (0.026\,K for 2t,
1.9\,Pa for msl, 4.6\,m$^2$\,s$^{-2}$ for z500 in bf16).

The consequences show directly in the forecasts. Aurora
converted as a whole to fp16 returns no finite value: every field is
infinite or undefined from the first 6-hour step on (three dates tested).
Converted as a whole to bf16, Aurora's forecasts remain finite but
their RMSE rises by up to 167\%, against at most 25\% with the network in
bf16 and the state in fp32 (Figure~\ref{fig:rmse-grid-naive} in
Appendix~\ref{sec:appendix-bf16}).

Keeping the state, its normalisation and its
denormalisation in fp32, and rounding only the normalised fields as they
enter the network, avoids these problems. It costs one fp32 copy of the
state -- for Aurora, the largest, 84 fields (with insolation) at two times, about 0.7\,GB,
6\% of its peak memory in fp16 -- and a few element-wise operations per
step. All 16-bit runs in the rest of this paper are built this way
(\S\ref{sec:quantisation}). Related considerations underlie the choice
of Aurora's developers to keep the computations before and after the
backbone in fp32 for its air-pollution extension
\citep{bodnar2024aurora}, and the practice in reduced-precision modelling of
rescaling the equations to fit the range of 16-bit numbers
\citep{klower2022fluid}.

\subsection{fp16 versus fp32}
\label{sec:results-fp16}

\begin{figure}[tbp]
\centering
\includegraphics[width=\textwidth]{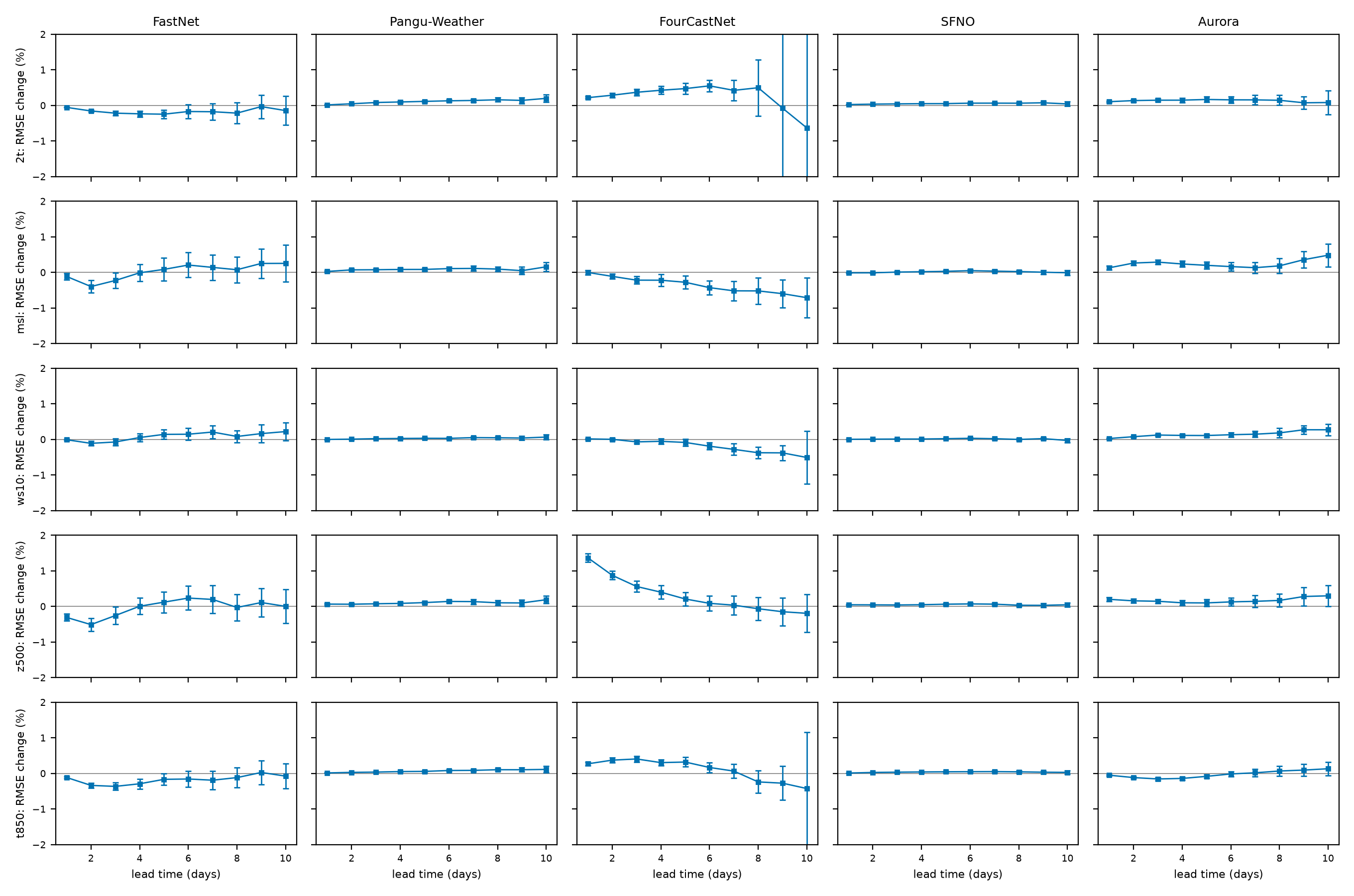}
\caption{Change in global RMSE against ERA5 of the fp16 forecasts relative
to fp32, $(\mathrm{fp16}-\mathrm{fp32})/\mathrm{fp32}$ of the mean over
the 52 dates of 2024, at days 1--10, with 95\% confidence intervals from
the paired per-date differences. Columns: models in order of size; rows:
variables. All panels show $-2$ to $+2\%$; FourCastNet's intervals for 2t
at days 9--10 and t850 at day 10 extend beyond this range.}
\label{fig:fp16-rmse}
\end{figure}

Figure~\ref{fig:fp16-rmse} shows the change in global RMSE
when the network runs in fp16. For FastNet, the changes lie between
$-0.5$ and $+0.25\%$; for 2t and t850, the fp16 forecasts are slightly more
accurate at almost every lead time. For Pangu-Weather, the RMSE increases by
up to 0.2\%, growing with lead time; the increase is small but
significant at the 5\% level for 45 of the 50 combinations of variable
and lead time. For FourCastNet, the changes lie between $-0.7$ and
$+1.4\%$: the largest increase is for z500 at day 1, and by day 10 the
fp16 forecasts are slightly more accurate than fp32 for every variable
($-0.2$ to $-0.7\%$). fp16 adds no instability: FourCastNet's largest 2t errors at day 10
fall on the same few dates in fp32 and fp16 (up to 19 and 15\,K, against a
median of 3.3\,K), and the largest ratio of fp16 to fp32 RMSE on a single
date is 1.3 (2t). For SFNO, the changes are the smallest of all
models, between $-0.03$ and $+0.07\%$, though still significant for 32 of
the 50 combinations. For Aurora, one of the two models most affected by bf16 (\S\ref{sec:results-bf16}), the changes lie between $-0.16$ and $+0.48\%$. The MAE behaves the same (not shown), with slightly larger
percentage changes (at most $+1.8\%$, for FourCastNet). Over the three regions the changes are of similar size, within about
$\pm2\%$ (SFNO: $\pm0.2\%$), except for FourCastNet over South Asia (up to $+5\%$)
(Table~\ref{tab:fp16-areas}). At the five locations, each represented by
only nine grid points, the changes are larger -- up to $\pm2\%$ for
Pangu-Weather and SFNO and up to about $\pm10\%$ for the other models --
but significant for few more combinations than expected by chance.

\begin{figure}[tbp]
\centering
\includegraphics[width=\textwidth]{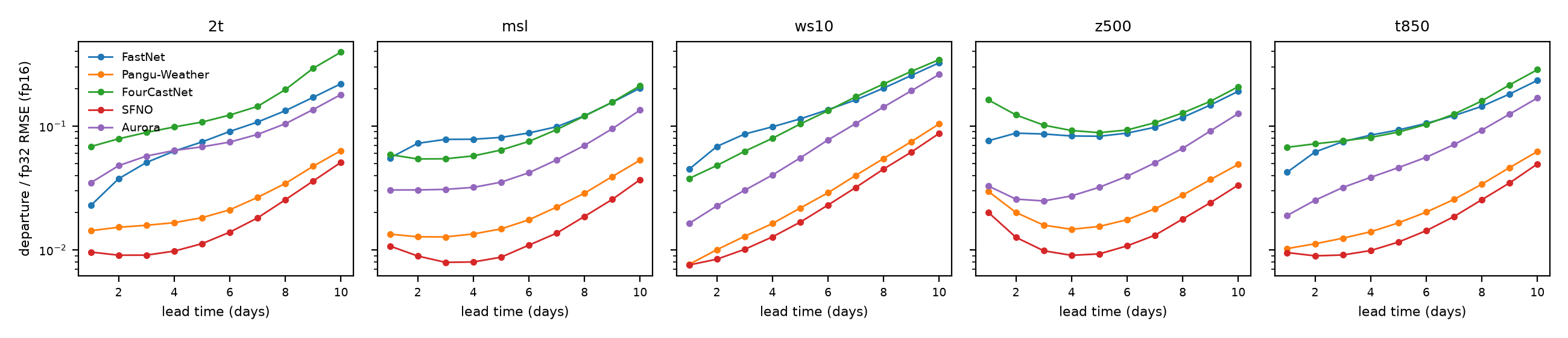}
\caption{Departure of the fp16 forecasts from the fp32 forecasts (RMS
difference, global) relative to the RMSE of the fp32 forecasts against
ERA5, mean over the 52 dates, at days 1--10.}
\label{fig:fp16-departure}
\end{figure}

Although their skill hardly changes, the fp16 forecasts drift
away from the fp32 forecasts, by day 10 to as much as 40\% of the forecast
error; the two behave like two equally skilful members of an ensemble
rather than a forecast and a degraded copy. Rounding in the network
perturbs every step, and the perturbations grow, as any small
perturbation of a chaotic system does (Figure~\ref{fig:fp16-departure}). For Pangu-Weather, the departure is
1--3\% of the fp32 error at day 1 and 5--10\% at day 10; for SFNO, 1--2\%
and 3--9\%; for Aurora, 2--4\% and 13--26\%; for FastNet,
2--8\% and 19--32\%; for FourCastNet, 4--16\% and 21--40\%. A departure of this size need not degrade the RMSE: two forecasts have
equal RMSE when the departure is weakly negatively correlated with the
error, with a correlation of minus half the relative departure (for a
departure of 40\% of the error, $-0.2$).

\begin{figure}[tbp]
\centering
\includegraphics[width=\textwidth]{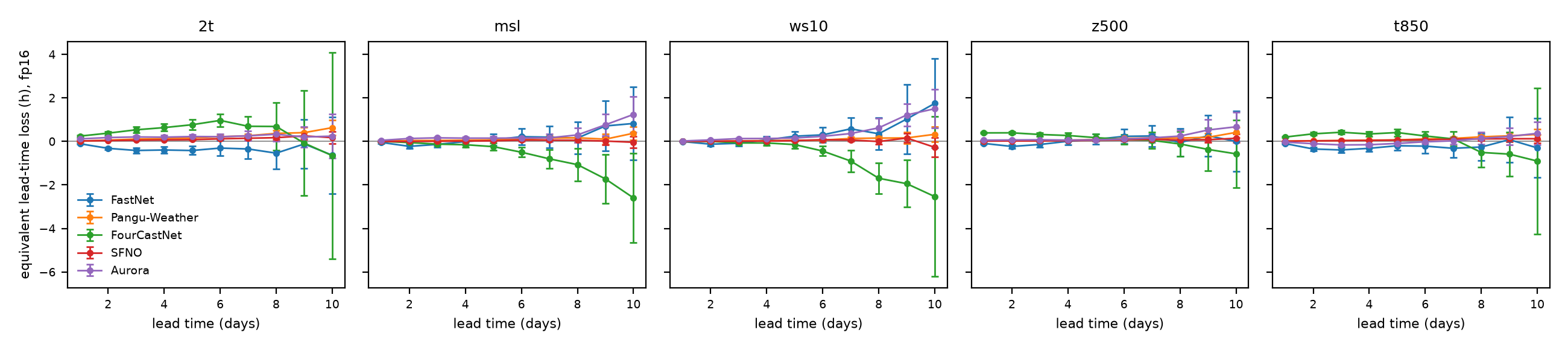}
\caption{The change in global RMSE of the fp16 forecasts expressed as an
equivalent change in lead time: the time over which the fp32 RMSE grows by
the same amount (positive: fp16 worse), with the 95\% interval of the
paired change; 52 dates in 2024.}
\label{fig:fp16-lead-time}
\end{figure}

To judge whether the RMSE changes between fp32 and fp16 matter in
practice, we express them as lead time. If the fp16 RMSE at lead time $t$
exceeds the fp32 RMSE by $\Delta$, the fp16 forecast is as accurate as the
fp32 forecast at the slightly longer lead time $t+\tau$, where the fp32
RMSE has grown by $\Delta$; for small $\tau$,
$\tau \approx \Delta / (\mathrm{d}\,\mathrm{RMSE}/\mathrm{d}t)$, with the
growth rate of the fp32 RMSE estimated from differences between daily
lead times. $\tau$ is the lead time the fp16 forecast loses (gains, if
negative; Figure~\ref{fig:fp16-lead-time}). Up to day 8, the fp16 forecasts lose
less than one hour of lead time for every model and variable, and often
gain a little; at day 10 the largest losses are 1.7\,h (FastNet) and 1.5\,h (Aurora), both for 10\,m wind speed. SFNO loses at most 0.3\,h at any lead time. The intervals widen at the longest lead times, where the error
grows more slowly and the conversion becomes less precise. For comparison,
forecasts of the leading global centres have gained about a day of lead
time per decade in the medium range \citep{bauer2015quiet}, or a few
hours per year.

\begin{table}[tbp]
\centering
\small
\caption{Inference cost on one RTX 5080: mean time per model step (first step of each forecast untimed, at least 60\,s of timed steps; FourCastNet, FastNet, SFNO and Aurora step 6\,h, Pangu-Weather 24\,h), speed-up over fp32, time for a 10-day forecast, peak allocated GPU memory, and energy for a 10-day forecast (mean GPU power within the timed steps, sampled every 50\,ms, times their duration). fp16: network in fp16, state in fp32. With the network in bf16 instead, time per step, peak memory and energy differ from fp16 by at most 2\%, 1\% and 3.4\%.}
\label{tab:benchmark}
\begin{tabular}{llrrrrr}
\toprule
Model & Precision & s/step & speed-up & s/10\,d & GB & Wh/10\,d \\
\midrule
FastNet & fp32 & 0.294 & 1.0 & 12 & 2.2 & 1.07 \\
 & fp16 & 0.105 & 2.8 & 4 & 1.2 & 0.29 \\
\midrule
Pangu-Weather & fp32 & 0.785 & 1.0 & 8 & 9.5 & 0.69 \\
 & fp16 & 0.371 & 2.1 & 4 & 5.2 & 0.28 \\
\midrule
FourCastNet & fp32 & 0.115 & 1.0 & 5 & 1.2 & 0.38 \\
 & fp16 & 0.057 & 2.0 & 2 & 0.9 & 0.14 \\
\midrule
SFNO & fp32 & 0.531 & 1.0 & 21 & 15.4 & 1.53 \\
 & fp16 & 0.400 & 1.3 & 16 & 12.9 & 1.07 \\
\midrule
Aurora$^\dagger$ & fp32 & 5.378 & 1.0 & 215 & 14.3 & 18.27 \\
 & fp16 & 1.517 & 3.5 & 61 & 11.6 & 5.50 \\
\bottomrule
\end{tabular}\\[2pt]
{\scriptsize $^\dagger$~In fp32, Aurora fits in 16\,GB only with its backbone weights streamed from CPU memory one block at a time. This slows its fp32 runs and keeps their peak memory down, so that its speed-up is partly due to the streaming, its memory saving is understated, and its energy comparison may be affected.}
\end{table}

Table~\ref{tab:benchmark} shows the cost of a forecast. fp16
networks run 2.0 (FourCastNet) to 3.5 (Aurora) times faster than fp32
(bf16 networks at the same speed, within 2\%), except for SFNO, which gains only a factor of
1.3; for Aurora, part of the gain comes from not having to stream its
weights from CPU memory, which fp32 requires on a 16\,GB GPU. SFNO's
small gain is plausibly due to its spherical harmonic transforms, which
stay in fp32, and to our application of its complex filters with four
real products (\S\ref{sec:models}); we have not profiled it. Peak memory
falls by about half for Pangu-Weather and FastNet, by 29\% for
FourCastNet, by a fifth for Aurora and by a sixth for SFNO. A 10-day
forecast uses 2.5 (Pangu-Weather) to 3.7 (FastNet) times less energy in
fp16 than in fp32, and SFNO 1.4 times less. The GPU draws less power in
fp16 for the four smaller models (by 7--27\%) but slightly more for
Aurora (by 7\%), so that the energy saving comes mostly from the shorter
run time.

\begin{table}[tbp]
\centering
\scriptsize
\setlength{\tabcolsep}{2.5pt}
\caption{Change in RMSE (\%) of the fp16 forecasts relative to fp32, $(\mathrm{fp16}-\mathrm{fp32})/\mathrm{fp32}$, at days 1, 5, 10, per area; 52 dates in 2024. Bold: significant at the 5\% level (paired $t$-test).}
\label{tab:fp16-areas}
\begin{tabular}{lrrrrrrrrrrrrrrr}
\toprule
 & \multicolumn{3}{c}{2t} & \multicolumn{3}{c}{msl} & \multicolumn{3}{c}{ws10} & \multicolumn{3}{c}{z500} & \multicolumn{3}{c}{t850} \\
 & d1 & d5 & d10 & d1 & d5 & d10 & d1 & d5 & d10 & d1 & d5 & d10 & d1 & d5 & d10 \\
\midrule
\multicolumn{16}{l}{\textit{FastNet}} \\
Global & \textbf{-0.1} & \textbf{-0.2} & -0.1 & \textbf{-0.1} & +0.1 & +0.3 & 0.0 & \textbf{+0.1} & +0.2 & \textbf{-0.3} & +0.1 & 0.0 & \textbf{-0.1} & \textbf{-0.2} & -0.1 \\
UK/Europe & \textbf{-0.1} & \textbf{-0.7} & -1.0 & \textbf{+0.5} & \textbf{-1.5} & -0.8 & 0.0 & -0.3 & +0.6 & \textbf{-0.5} & \textbf{-1.2} & -1.7 & 0.0 & \textbf{-1.1} & -1.0 \\
North America & 0.0 & \textbf{-0.5} & -0.5 & -0.2 & -0.3 & -0.6 & -0.1 & +0.2 & -0.8 & \textbf{-0.8} & -0.5 & \textbf{-1.4} & -0.1 & \textbf{-0.6} & -0.6 \\
South Asia & \textbf{0.0} & \textbf{-0.2} & -0.2 & \textbf{-0.2} & -0.2 & -0.6 & 0.0 & \textbf{-0.2} & -0.3 & \textbf{-0.3} & \textbf{-1.1} & +0.1 & 0.0 & -0.1 & +0.3 \\
London & -0.4 & -2.4 & -1.6 & +2.8 & -1.6 & +3.0 & -0.2 & -2.0 & -2.9 & -2.2 & -2.6 & +1.4 & -0.5 & -3.1 & +0.4 \\
Delhi & -0.2 & -0.2 & \textbf{-5.0} & \textbf{-1.1} & +0.5 & +3.3 & 0.0 & -0.3 & +5.7 & \textbf{-0.7} & \textbf{-2.3} & +2.6 & -0.1 & 0.0 & +0.5 \\
São Paulo & \textbf{+0.7} & -2.0 & +3.4 & +0.5 & +0.2 & +2.8 & \textbf{-0.8} & -1.7 & -4.4 & -1.5 & +1.2 & -1.8 & -0.6 & -1.9 & +0.9 \\
Bay Area & +0.1 & \textbf{-1.3} & -0.8 & -1.4 & -1.0 & +4.0 & \textbf{+1.5} & -1.3 & -2.6 & -0.5 & +0.5 & -2.4 & +0.8 & -0.1 & -3.0 \\
Beijing & -0.2 & -0.3 & -1.1 & \textbf{-1.2} & \textbf{-2.5} & +2.2 & 0.0 & +0.5 & +10.6 & \textbf{-1.8} & -1.4 & -2.1 & 0.0 & -0.3 & +4.6 \\
\midrule
\multicolumn{16}{l}{\textit{Pangu-Weather}} \\
Global & \textbf{0.0} & \textbf{+0.1} & \textbf{+0.2} & \textbf{0.0} & \textbf{+0.1} & \textbf{+0.2} & 0.0 & \textbf{0.0} & +0.1 & \textbf{+0.1} & \textbf{+0.1} & \textbf{+0.2} & \textbf{0.0} & \textbf{+0.1} & \textbf{+0.1} \\
UK/Europe & 0.0 & 0.0 & +0.1 & +0.1 & 0.0 & -0.1 & 0.0 & 0.0 & 0.0 & 0.0 & +0.1 & 0.0 & \textbf{0.0} & +0.1 & 0.0 \\
North America & \textbf{0.0} & 0.0 & -0.1 & 0.0 & \textbf{+0.1} & -0.3 & \textbf{0.0} & \textbf{+0.1} & 0.0 & 0.0 & \textbf{+0.1} & -0.3 & 0.0 & +0.1 & -0.2 \\
South Asia & \textbf{0.0} & \textbf{+0.1} & \textbf{+0.2} & 0.0 & -0.1 & +0.5 & 0.0 & +0.1 & \textbf{+0.3} & \textbf{+0.1} & 0.0 & +0.2 & 0.0 & 0.0 & 0.0 \\
London & 0.0 & +0.2 & -0.8 & 0.0 & 0.0 & -0.4 & 0.0 & -0.1 & -2.0 & -0.1 & 0.0 & -0.7 & 0.0 & +0.1 & -0.4 \\
Delhi & -0.1 & 0.0 & +0.3 & -0.1 & -0.1 & +0.9 & -0.1 & 0.0 & +0.5 & +0.3 & +0.2 & +0.2 & 0.0 & 0.0 & +0.1 \\
São Paulo & -0.1 & -0.1 & -0.3 & 0.0 & -0.3 & -0.2 & 0.0 & 0.0 & 0.0 & 0.0 & -0.1 & +0.4 & -0.1 & \textbf{+0.4} & -0.6 \\
Bay Area & \textbf{+0.2} & \textbf{+0.4} & +0.8 & +0.3 & +0.3 & -0.4 & +0.1 & +0.4 & +0.7 & -0.6 & +0.3 & +1.1 & +0.1 & +0.2 & +0.2 \\
Beijing & 0.0 & +0.1 & +0.3 & -0.1 & +0.3 & -0.2 & -0.1 & -0.1 & +2.0 & +0.1 & -0.1 & +0.4 & \textbf{+0.2} & 0.0 & -0.7 \\
\midrule
\multicolumn{16}{l}{\textit{FourCastNet}} \\
Global & \textbf{+0.2} & \textbf{+0.5} & -0.6 & 0.0 & \textbf{-0.3} & \textbf{-0.7} & 0.0 & -0.1 & -0.5 & \textbf{+1.4} & \textbf{+0.2} & -0.2 & \textbf{+0.3} & \textbf{+0.3} & -0.4 \\
UK/Europe & -0.1 & +0.1 & -0.4 & +0.2 & -0.2 & -0.6 & 0.0 & -0.1 & -0.5 & \textbf{+1.5} & +0.1 & -0.9 & \textbf{+0.4} & +0.2 & -1.2 \\
North America & 0.0 & -0.4 & -0.1 & 0.0 & -0.5 & +0.1 & 0.0 & 0.0 & +0.2 & \textbf{+1.1} & -0.3 & +1.0 & \textbf{+0.2} & -0.2 & -0.2 \\
South Asia & \textbf{+0.2} & 0.0 & -0.2 & +0.2 & -0.2 & -0.5 & 0.0 & \textbf{-0.5} & -0.2 & \textbf{+2.0} & \textbf{+4.2} & \textbf{+3.0} & +0.2 & -0.3 & -0.5 \\
London & -0.2 & -2.5 & -0.3 & -1.1 & +0.3 & -1.4 & -0.5 & -0.1 & +1.5 & +1.9 & +1.4 & -2.3 & \textbf{+2.3} & \textbf{+3.7} & -0.5 \\
Delhi & -0.3 & -1.2 & -1.0 & -0.1 & -0.9 & -0.2 & \textbf{-0.8} & -0.1 & +0.2 & +3.4 & +3.9 & 0.0 & 0.0 & -0.9 & +3.5 \\
São Paulo & -1.2 & +0.5 & +3.0 & +1.1 & 0.0 & \textbf{+5.3} & -0.2 & -0.2 & +2.0 & +0.8 & +2.7 & +3.9 & +0.8 & -1.0 & +2.4 \\
Bay Area & +0.2 & +1.4 & +1.8 & 0.0 & +0.7 & +0.9 & -0.3 & -0.9 & -3.9 & \textbf{+5.6} & +0.7 & -0.2 & -0.8 & -0.3 & -2.5 \\
Beijing & -0.6 & \textbf{-3.6} & -4.1 & \textbf{+2.0} & +1.7 & -4.9 & +0.1 & -0.9 & -0.2 & \textbf{+5.5} & +1.4 & +3.0 & +0.5 & \textbf{-2.8} & +0.5 \\
\midrule
\multicolumn{16}{l}{\textit{SFNO}} \\
Global & \textbf{0.0} & \textbf{0.0} & 0.0 & \textbf{0.0} & \textbf{0.0} & 0.0 & 0.0 & \textbf{0.0} & 0.0 & \textbf{0.0} & \textbf{+0.1} & 0.0 & \textbf{0.0} & \textbf{0.0} & 0.0 \\
UK/Europe & 0.0 & 0.0 & 0.0 & \textbf{-0.1} & 0.0 & 0.0 & 0.0 & 0.0 & +0.1 & 0.0 & +0.1 & 0.0 & 0.0 & 0.0 & -0.1 \\
North America & 0.0 & 0.0 & 0.0 & 0.0 & \textbf{-0.1} & 0.0 & \textbf{0.0} & 0.0 & -0.1 & 0.0 & 0.0 & +0.1 & 0.0 & 0.0 & +0.1 \\
South Asia & 0.0 & 0.0 & 0.0 & 0.0 & 0.0 & 0.0 & \textbf{0.0} & 0.0 & 0.0 & 0.0 & 0.0 & +0.1 & 0.0 & 0.0 & +0.1 \\
London & -0.1 & 0.0 & +0.1 & -0.1 & 0.0 & \textbf{+0.9} & -0.1 & +0.1 & -0.5 & 0.0 & +0.1 & +0.5 & -0.2 & 0.0 & +0.4 \\
Delhi & \textbf{+0.2} & 0.0 & 0.0 & +0.1 & +0.2 & +0.1 & 0.0 & +0.1 & -0.6 & -0.3 & +0.1 & -0.2 & +0.2 & -0.1 & -0.1 \\
São Paulo & 0.0 & 0.0 & -0.2 & -0.1 & +0.1 & -0.6 & -0.1 & -0.2 & -0.7 & 0.0 & 0.0 & +0.1 & +0.2 & 0.0 & -0.3 \\
Bay Area & 0.0 & 0.0 & +0.2 & +0.1 & +0.1 & +0.5 & 0.0 & 0.0 & -0.5 & 0.0 & 0.0 & +0.1 & +0.1 & 0.0 & +0.1 \\
Beijing & 0.0 & -0.1 & -0.7 & 0.0 & +0.2 & -1.2 & 0.0 & 0.0 & +0.5 & +0.2 & +0.2 & +0.3 & 0.0 & 0.0 & -0.9 \\
\midrule
\multicolumn{16}{l}{\textit{Aurora}} \\
Global & \textbf{+0.1} & \textbf{+0.2} & +0.1 & \textbf{+0.1} & \textbf{+0.2} & \textbf{+0.5} & \textbf{0.0} & \textbf{+0.1} & \textbf{+0.3} & \textbf{+0.2} & \textbf{+0.1} & \textbf{+0.3} & \textbf{-0.1} & \textbf{-0.1} & +0.1 \\
UK/Europe & 0.0 & +0.2 & \textbf{+0.8} & \textbf{+0.2} & +0.1 & +0.3 & \textbf{0.0} & +0.1 & 0.0 & 0.0 & 0.0 & +0.9 & 0.0 & 0.0 & +0.6 \\
North America & \textbf{0.0} & \textbf{+0.2} & +0.9 & \textbf{+0.2} & \textbf{+0.3} & +0.3 & \textbf{0.0} & +0.1 & \textbf{-0.6} & \textbf{+0.2} & \textbf{+0.3} & +0.3 & \textbf{-0.1} & +0.1 & +0.6 \\
South Asia & 0.0 & +0.2 & \textbf{+0.7} & +0.1 & \textbf{+1.0} & \textbf{+1.9} & \textbf{+0.1} & \textbf{+0.6} & \textbf{+1.1} & \textbf{+0.5} & \textbf{-0.9} & -0.3 & 0.0 & +0.2 & +0.4 \\
London & +0.2 & \textbf{+1.5} & +3.5 & 0.0 & +0.1 & +1.8 & 0.0 & 0.0 & +2.8 & -0.6 & \textbf{+1.0} & +3.0 & -0.2 & +0.7 & -1.1 \\
Delhi & -0.1 & \textbf{-1.5} & -0.7 & +0.9 & -0.6 & +2.9 & +0.1 & -0.8 & -0.2 & -0.2 & \textbf{-2.3} & +1.1 & -0.4 & \textbf{-1.5} & +2.2 \\
São Paulo & \textbf{-0.3} & +0.1 & +1.0 & -0.3 & -0.5 & +1.2 & 0.0 & +0.5 & -0.2 & +0.1 & \textbf{-2.3} & +1.2 & -0.3 & +0.6 & -0.3 \\
Bay Area & \textbf{+0.5} & \textbf{+2.5} & +3.4 & 0.0 & 0.0 & -0.8 & 0.0 & -0.7 & +7.3 & -0.1 & -0.1 & +0.2 & +0.4 & -0.7 & -2.4 \\
Beijing & -0.2 & +0.2 & +0.4 & +0.3 & +1.0 & -0.1 & -0.1 & -0.7 & +1.2 & -0.7 & +0.6 & -0.4 & 0.0 & +0.6 & -1.3 \\
\bottomrule
\end{tabular}
\end{table}

\subsection{Where fp16 and fp32 forecasts differ}
\label{sec:results-where}

\begin{figure}[tbp]
\centering
\includegraphics[width=\textwidth]{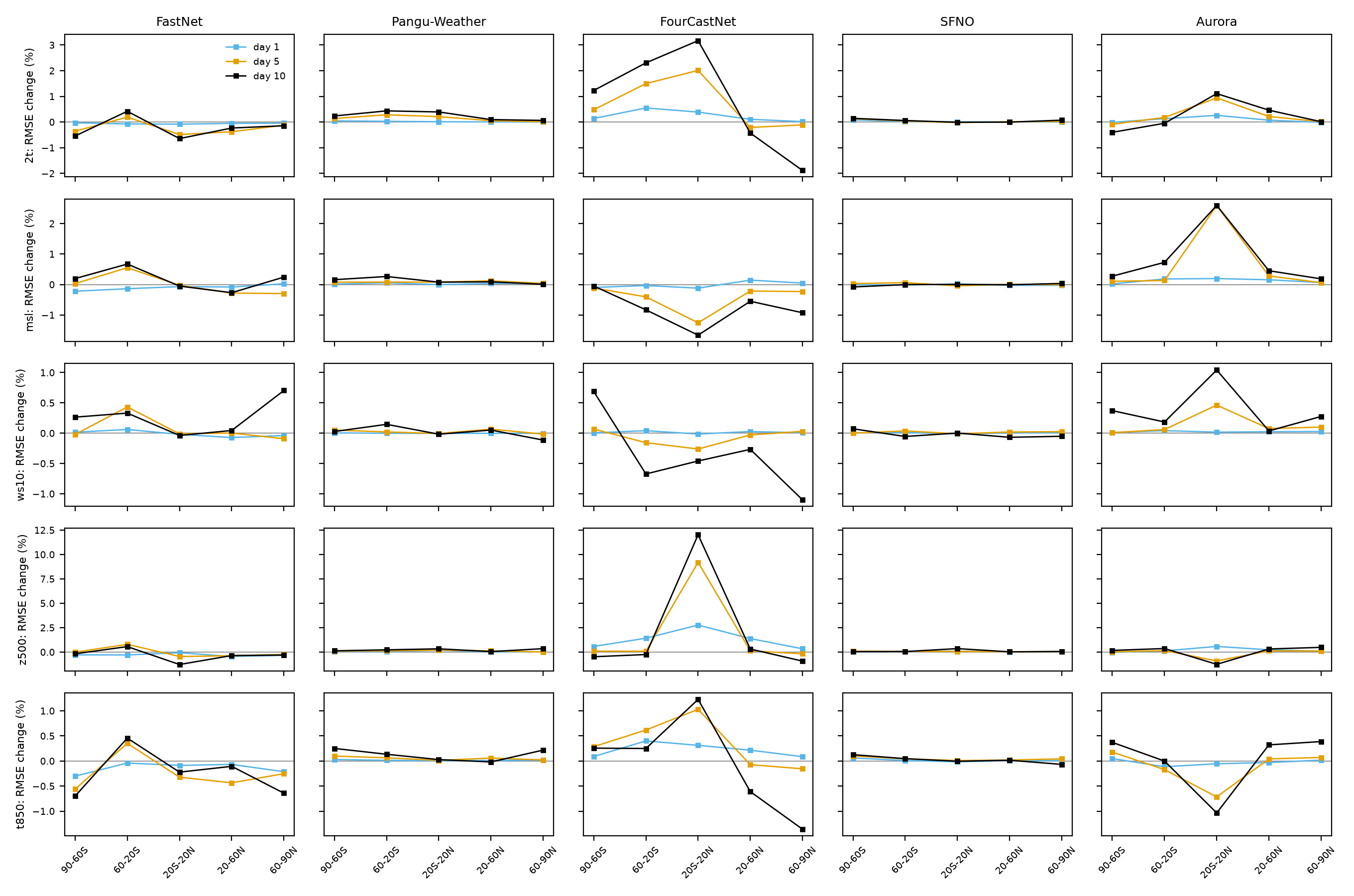}
\caption{Change in RMSE against ERA5 of the fp16 forecasts relative to fp32,
by latitude band, at days 1, 5 and 10; 52 dates in 2024.}
\label{fig:fp16-bands}
\end{figure}

Figure~\ref{fig:fp16-bands} splits the change in RMSE by
latitude band. For FastNet, Pangu-Weather, SFNO and Aurora, it stays within $\pm1.3\%$ in every band (SFNO: within $\pm0.4\%$), except for Aurora's tropical msl (up to $+2.6\%$ at days 5 and 10). FourCastNet's near-zero global change hides opposite changes
in different bands: between 20$^\circ$S and 20$^\circ$N, its fp16 z500
RMSE is 2.8\% higher than fp32 at day 1 and 12\% higher at day 10, and
its 2t and t850 RMSE up to 3\% and 1\% higher, while at high latitudes
fp16 is up to 2\% more accurate. The tropical z500 error is small in
absolute terms (fp32 RMSE 44 and 155\,m$^2$\,s$^{-2}$ at days 1 and 10,
against 66 and 907\,m$^2$\,s$^{-2}$ globally), so that the change there
contributes little to the global score.

\begin{figure}[tbp]
\centering
\includegraphics[width=\textwidth]{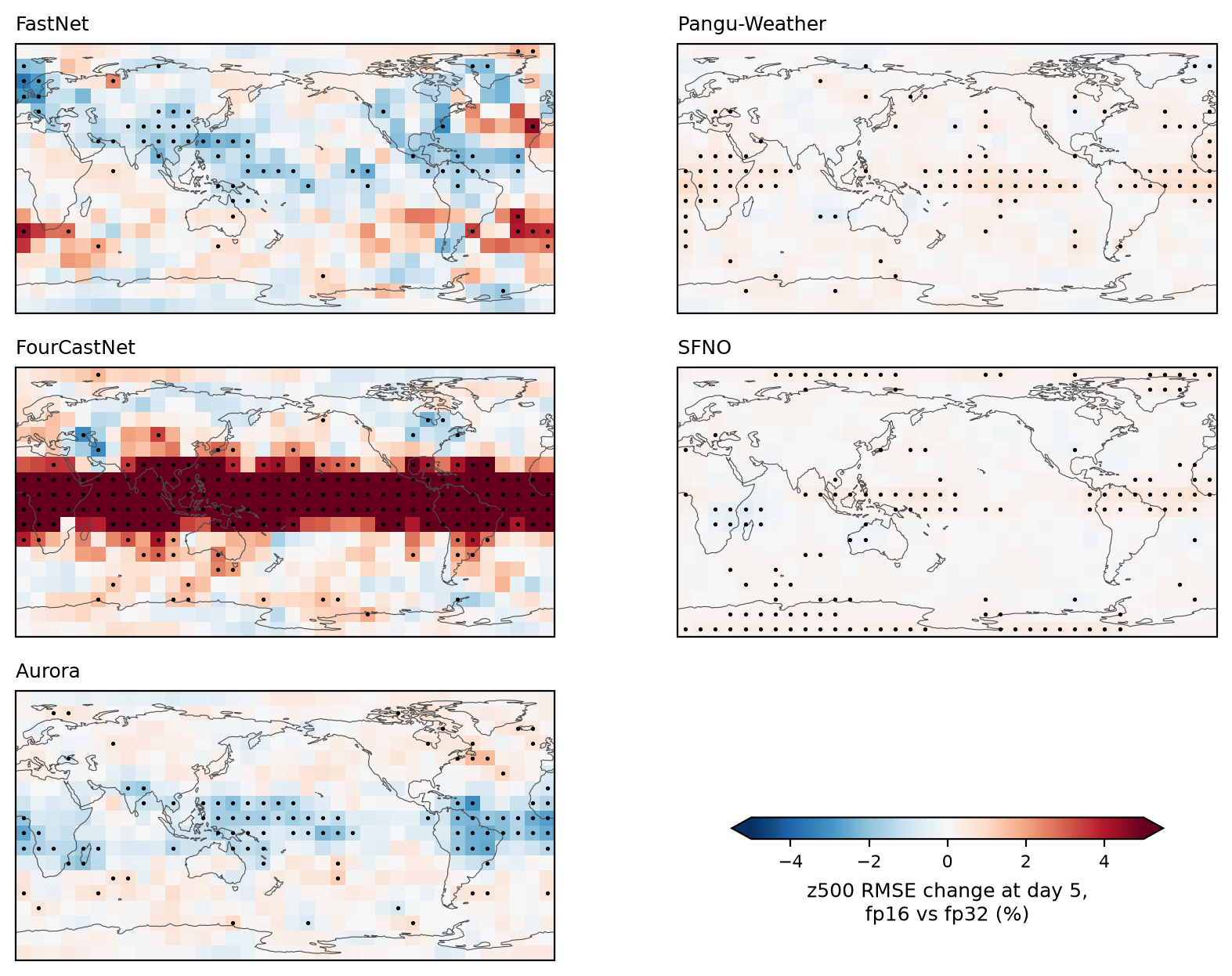}
\caption{Change in z500 RMSE against ERA5 of the fp16 forecasts relative
to fp32 at day 5, on $10^\circ\times10^\circ$ cells; 52 dates in 2024.
Dots: cells where the paired $t$-test across dates is significant at the
5\% level (with 648 cells, about 5\% of them would be by chance). The
colour scale ends at $\pm5\%$; FourCastNet's tropical cells exceed it.}
\label{fig:fp16-cells}
\end{figure}

Figure~\ref{fig:fp16-cells} maps the change in z500 RMSE at
day 5 on the $10^\circ$ cells. For Pangu-Weather and SFNO it stays within
$\pm1\%$ in every cell, and for FastNet and Aurora within $\pm2\%$ in 91\%
and 97\% of the cells. FastNet's changes form a patchy pattern of either
sign, up to $\pm4\%$; Aurora's fp16 forecasts are slightly more accurate
than fp32, by 1--3\%, over tropical South America, Africa and the western
Pacific. FourCastNet's fp16 penalty fills the tropics: its tropical cells
are 11\% worse on the median and up to 18\%, the cause of its tropical
penalty in Figure~\ref{fig:fp16-bands}. More cells than the 5\% expected
by chance are significant (11--30\%), but for Pangu-Weather and SFNO the
significant changes are below 1\%: they are detectable because the paired
design is so sensitive, not because they are large.

\begin{figure}[tbp]
\centering
\includegraphics[width=\textwidth]{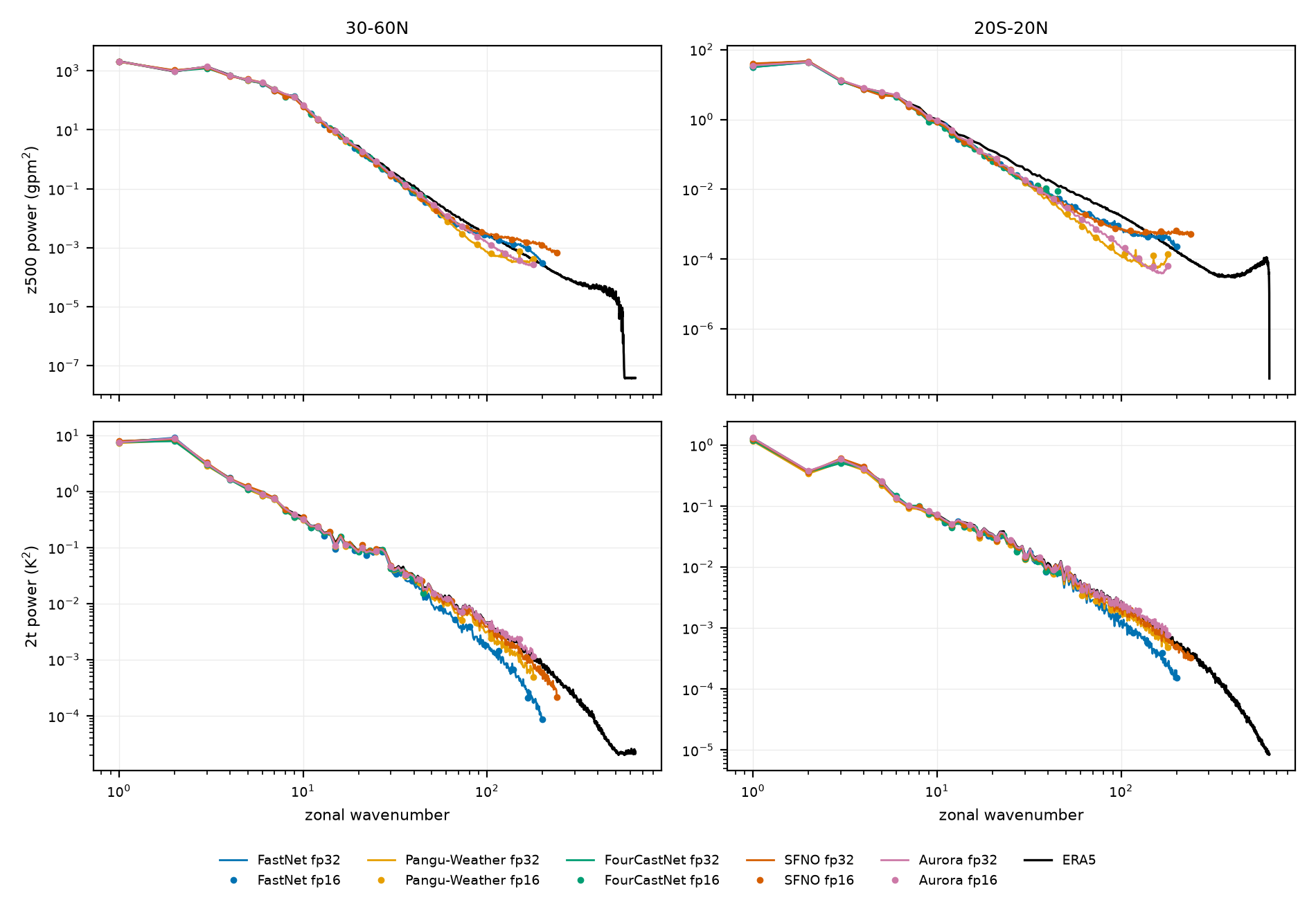}
\caption{Zonal power spectra of z500 (top, geopotential height) and 2t
(bottom) at day 5, averaged over the 52 dates and over 30--60$^\circ$N
(left) and 20$^\circ$S--20$^\circ$N (right): ERA5 (black), fp32 forecasts
(lines) and fp16 forecasts (dots, same colour), each model up to the
smallest scale it resolves (wavenumber 45 for FourCastNet, 179 for Aurora and Pangu-Weather, 200 for FastNet, 239 for SFNO).}
\label{fig:fp16-spectra}
\end{figure}

Figure~\ref{fig:fp16-spectra} compares the zonal power spectra
of the fp32 and fp16 forecasts with that of ERA5. At all scales the models
resolve, the fp16 spectra lie on the fp32 spectra for 2t, and for z500 in
FastNet and SFNO and in the mid-latitudes for all models. At the smallest scales,
the fp16 forecasts contain slightly more z500 variance: for Pangu-Weather
and Aurora, in the tropics at wavenumbers above about 100 (wavelengths
below 400\,km), where the RMS amplitude of these waves rises from 0.11 to
0.12--0.13\,gpm, against 0.34\,gpm in ERA5 -- the fp16 forecasts remain far
smoother than ERA5 there. For FourCastNet, fp16 adds tropical z500 variance from wavenumber about 25 (1600\,km at the equator) up to its limit of 45: the RMS amplitude of these waves rises from 0.73 to 0.81\,gpm, against 1.12\,gpm in ERA5, consistent with its higher tropical z500 error (Figure~\ref{fig:fp16-bands}); in the mid-latitudes its spectra are unchanged. Above wavenumber 45, which FourCastNet's spectral filters do not act on, its fp32 spectra lie 10--100 times above ERA5's, and fp16 adds further variance there (not shown). The added variance is consistent
with rounding noise in the network output, which has a flat spectrum and
is visible only where the forecast itself has little variance. In
absolute terms it amounts to a few tenths of a geopotential metre at most,
and does not smooth the forecasts.

\begin{figure}[tbp]
\centering
\includegraphics[width=\textwidth]{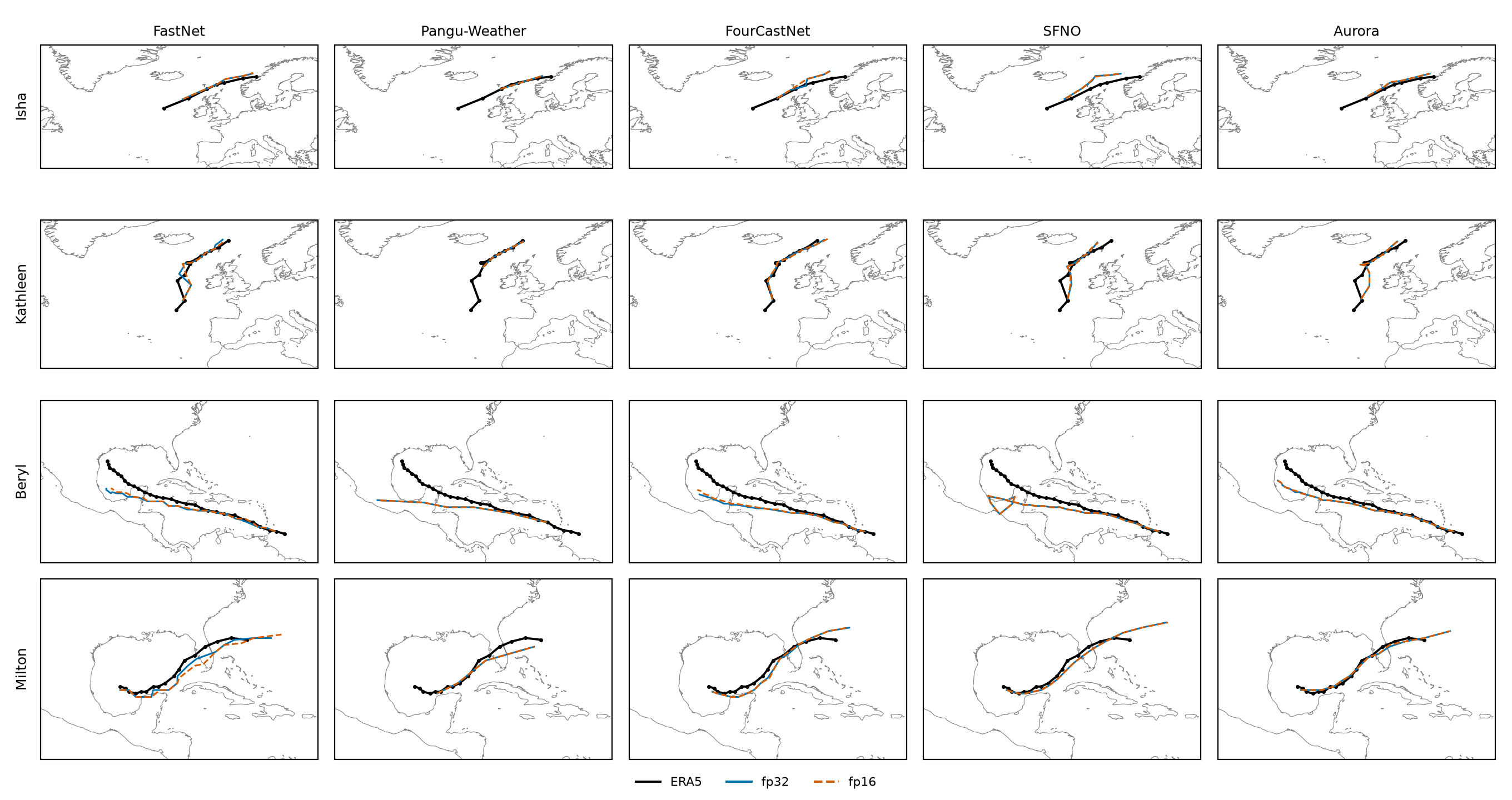}
\caption{Tracks of the storms Isha, Kathleen, Beryl and Milton (rows) in
ERA5 (black) and in the fp32 (solid) and fp16 (dashed) forecasts of each
model (columns), followed from the ERA5 position at the first tracked
lead time. Pangu-Weather's tracks have 24-hour steps.}
\label{fig:storm-tracks}
\end{figure}

Figure~\ref{fig:storm-tracks} shows the tracks of the four
storms. For Pangu-Weather, the fp16 and fp32 tracks are identical to
within a kilometre; for Aurora and SFNO, the mean position errors
differ by at most 3\,km. For FastNet and FourCastNet, the fp16 tracks depart from the
fp32 tracks later in the forecast, in either direction: the mean position
error changes by between $-29$ and $+34$\,km per storm, the largest change
being FastNet's Milton forecast (163 to 197\,km), which separates from
the fp32 track after landfall. The mean central-pressure error changes by
at most 0.7\,hPa for every model and storm. These differences are much
smaller than those between the models and ERA5 (mean position errors of
60--430\,km). Four cases cannot show systematic differences; they show
that fp16 does not visibly degrade the forecasts of these storms.

\begin{figure}[tbp]
\centering
\includegraphics[width=\textwidth]{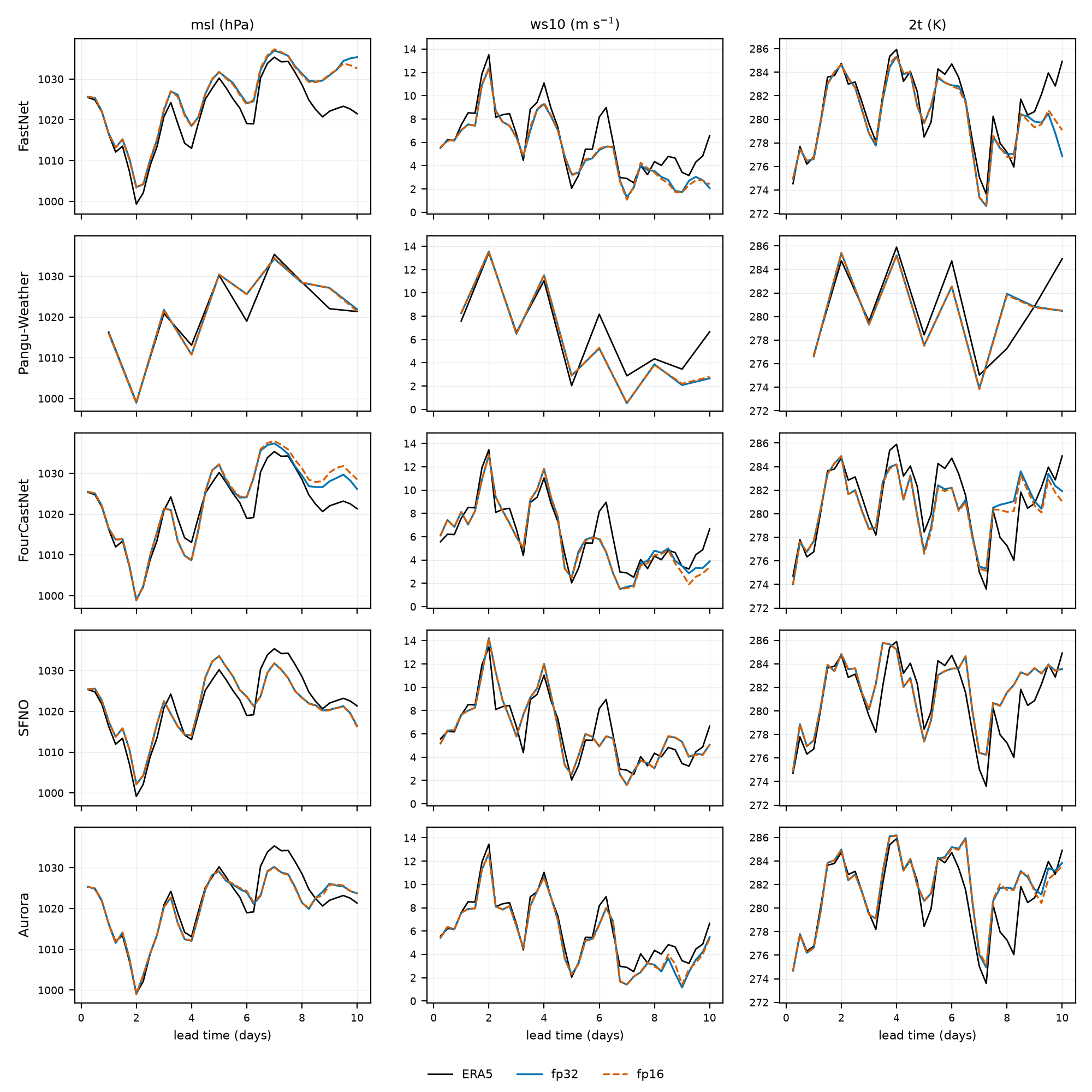}
\caption{Forecasts for London (mean over the $3\times3$ grid points
nearest to its centre; for FastNet, over the six its interpolated fields
cover) from 20 January 2024 00\,UTC, through Storms Isha (minimum on
22 January, day 2) and Jocelyn (24 January, day 4), which crossed the UK
in quick succession: mean sea-level pressure, 10\,m wind speed and 2\,m
temperature in ERA5 (black) and in the
fp32 (solid) and fp16 (dashed) forecasts of each model (rows).
Pangu-Weather's forecasts have 24-hour steps.}
\label{fig:location-series}
\end{figure}

Figure~\ref{fig:location-series} shows how the fp16 and fp32
forecasts diverge at a single location, London, in a forecast through
two storms in quick succession, Isha and Jocelyn, the strongest of which,
Isha, was the most significant wind storm in the UK for almost two years
\citep{metoffice2024storms}. Up to day 5, the two are indistinguishable at the scale of the
forecast: they differ by at most 0.5\,hPa, 0.3\,m\,s$^{-1}$ and 0.3\,K
(FastNet; Pangu-Weather, SFNO and Aurora by at most 0.15\,hPa,
0.03\,m\,s$^{-1}$ and 0.05\,K), against mean fp32 errors of 0.7--2.3\,hPa,
0.3--0.8\,m\,s$^{-1}$ and 0.4--1.0\,K over the same days. All models
capture the deepening to about 1000\,hPa at day 2 and the wind maximum
with it, and the second low two days later, in fp16 as in fp32: in this
high-impact situation the fp16 forecasts are as stable as the fp32
forecasts. From day 6, the fp16 forecasts of FastNet,
FourCastNet and Aurora drift away from fp32, by up to 2.8\,hPa,
1.0\,m\,s$^{-1}$ and 2.2\,K, in either direction and by amounts comparable
to or smaller than the forecasts' errors at these lead times (mean fp32
errors of 3.7--5.5\,hPa, 1.0--1.4\,m\,s$^{-1}$ and 1.6--1.7\,K); those of
Pangu-Weather and SFNO stay within 0.5\,hPa, 0.13\,m\,s$^{-1}$ and 0.11\,K
of fp32 throughout. This is one case; the location statistics over all 52
dates are in Table~\ref{tab:fp16-areas}.

\subsection{fp16 ensembles}
\label{sec:results-ensemble}

\begin{figure}[tbp]
\centering
\includegraphics[width=\textwidth]{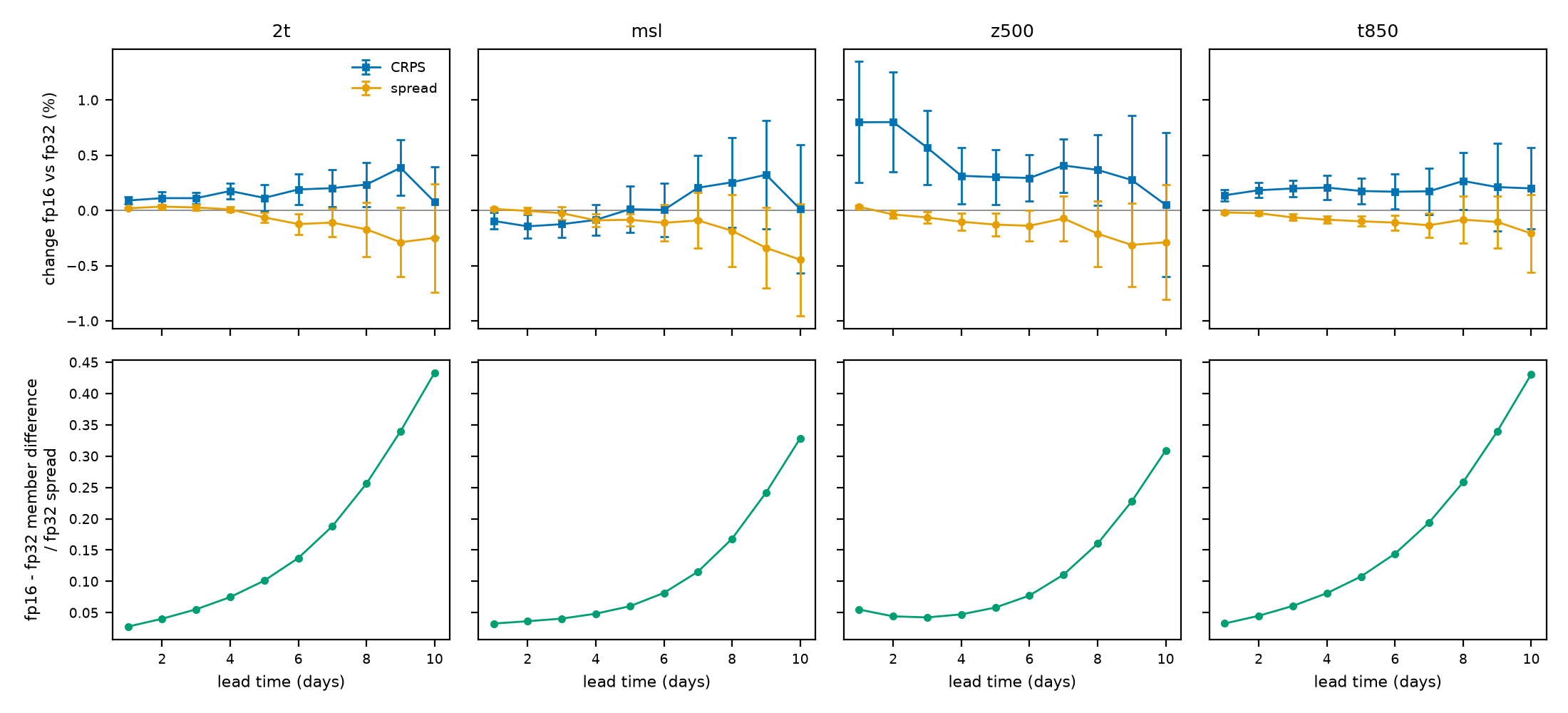}
\caption{Aurora ensemble (13 dates in 2024, 8 members): change of the fp16
ensemble relative to fp32 in global CRPS and spread, with 95\% intervals
from the paired per-date differences (top), and RMS difference between
each fp16 member and the fp32 member with the same noise, relative to the
fp32 spread (bottom).}
\label{fig:fp16-ensemble}
\end{figure}

For Aurora's ensemble checkpoint, fp16 changes the global CRPS
by at most $+0.8\%$ (z500 at day 1) and mostly by 0.0--0.3\%, and the
ensemble spread by less than 0.5\% (Figure~\ref{fig:fp16-ensemble}). The
fraction of the globe where ERA5 lies outside the range of the members
(0.21--0.38, depending on variable and lead time) is unchanged to within
0.01, so fp16 does not change the calibration of the ensemble. The
ensemble's CRPS is 0.56--0.76 times the MAE of Aurora's deterministic
forecasts on the same dates (2t and z500), the usual advantage of a
well-spread ensemble over a single forecast, and fp16 changes neither.

Each fp16 member departs from the fp32 member with the same
noise by 3--11\% of the ensemble spread up to day 5 and by 31--43\% at day
10 (Figure~\ref{fig:fp16-ensemble}, bottom). The rounding in fp16 thus
acts like a small additional perturbation, much smaller than the
perturbations the ensemble is designed to sample, and growing like them.

\subsection{Range headroom of fp16}
\label{sec:results-headroom}

\begin{figure}[tbp]
\centering
\includegraphics[width=0.75\textwidth]{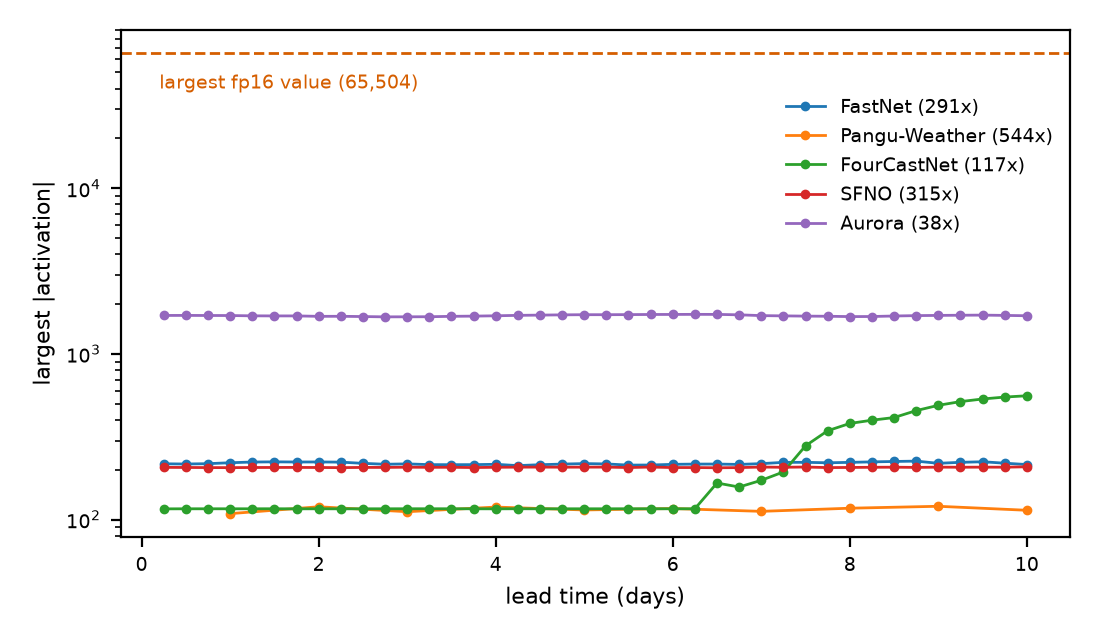}
\caption{Largest magnitude of any fp16 value returned by any module of
each model, per model step, over the 52 dates of 2024, against the largest
finite fp16 value (65,504, dashed). In parentheses: the ratio of 65,504 to
the largest value over all steps.}
\label{fig:fp16-headroom}
\end{figure}

fp16's limited numeric range is its main risk: a value above 65,504
anywhere in the network becomes infinite, and the forecast is lost. On the 52 dates, no module of
any model returned a non-finite value. The largest magnitudes
(Figure~\ref{fig:fp16-headroom}) are 120 (Pangu-Weather), 208 (SFNO),
225 (FastNet), 559 (FourCastNet) and 1,733 (Aurora), 544, 315, 291, 117
and 38 times below the limit. For four of the models they are constant
over the 10 days;
FourCastNet's grow from 116 to 559 after day 6, in its input layer, probably because its normalised state drifts outside the range it was trained on late in the forecast. Aurora's largest values, which set its 38-fold margin, occur in
an MLP of the first stage of its backbone, the component that also
dominates its fp16 departure for 2t (\S\ref{sec:results-sensitivity}).

Values smaller than $6.1\times10^{-5}$, fp16's smallest
normal number, keep fewer significant bits. Over all modules, the median share of such values is 0.008--0.06\%.
The exceptions are outputs that are naturally close to zero: attention
weights after the softmax in Pangu-Weather (up to 24\%), activations after
the GELU in FourCastNet and Aurora (up to 17\% and 9\%) and the output of
some normalisation layers in FastNet (11\%). Since these values are tiny,
their reduced precision changes the sums they enter by a negligible
amount.

\subsection{Where fp16 rounding enters the forecast}
\label{sec:results-sensitivity}

\begin{figure}[tbp]
\centering
\includegraphics[width=\textwidth]{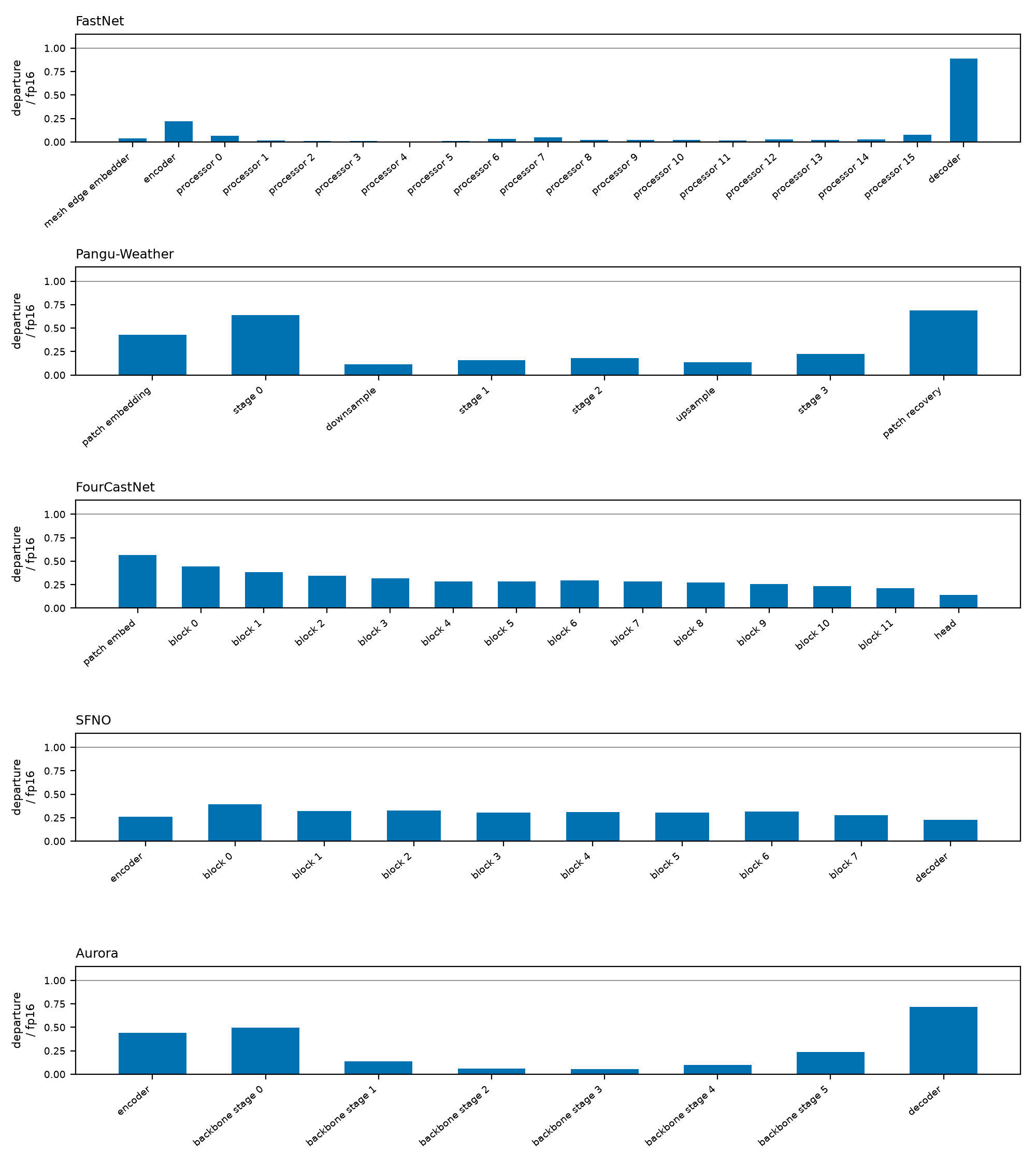}
\caption{Departure of the forecast from fp32 at 24\,h (z500, global, mean
over 13 dates) when only one component runs in fp16, relative to the
departure with the whole network in fp16. Components in execution order.}
\label{fig:fp16-oat}
\end{figure}

A common rule in reduced-precision inference is to keep the
first and last layers in higher precision \citep{choi2018pact,
bhattacharya2026ptq}. Figure~\ref{fig:fp16-oat} tests it. For Pangu-Weather, the
ends dominate: in fp16, the patch embedding, the first stage and the patch
recovery alone each produce 43--69\% of the full departure, the middle stages 11--22\%. Aurora behaves similarly: its encoder, first backbone stage and decoder dominate (44, 49 and 72\% of the z500 departure; for 2t, the first backbone stage alone 73\%). For FastNet, the last component
dominates: its decoder alone produces 89\% of the departure, whereas the
16 processor rounds contribute at most 8\% each. For FourCastNet,
sensitivity falls steadily with depth, from 57\% for the patch embedding
to 14\% for the output layer. For SFNO, every component contributes about
equally: each alone produces 23--39\% of the z500 departure. SFNO is also
the one model in which converting every component separately departs
much less than converting the whole network (48\% for z500, 68\% for 2t;
98--100\% for the other models). The difference lies in what stays in
fp32 when the components are converted separately: the activations
passed from one component to the next, which are converted back to fp32
at each component's boundary, and the $1\times1$ convolution on SFNO's
residual path, which carries the network's input directly to its output
and belongs to no component; we have not separated the two. The contributions of the components do
not add linearly but roughly in quadrature: the square root of the sum
of their squares is close to 1 (0.9--1.5 for the five models), as for
independent sources of noise.

Normalisation and softmax layers are often treated specially in
mixed-precision training, with their sums carried out in fp32
\citep{micikevicius2018mixed}. Keeping all normalisation layers, all
attention or spectral-filter layers, or all MLPs in fp32 changes the
z500 and 2t departures by at most 6\% for Aurora, Pangu-Weather, FourCastNet
and SFNO (whose filters stay complex fp32 numbers; up to 16\% for other
variables, Pangu-Weather's 10v with its attention layers in fp32), and keeping
FastNet's normalisation layers in fp32 by 5--10\%; only keeping all of
FastNet's linear layers in fp32 -- nearly the whole network -- reduces them
substantially (to 29--42\%; 19\% for 10v).

\begin{figure}[tbp]
\centering
\includegraphics[width=0.75\textwidth]{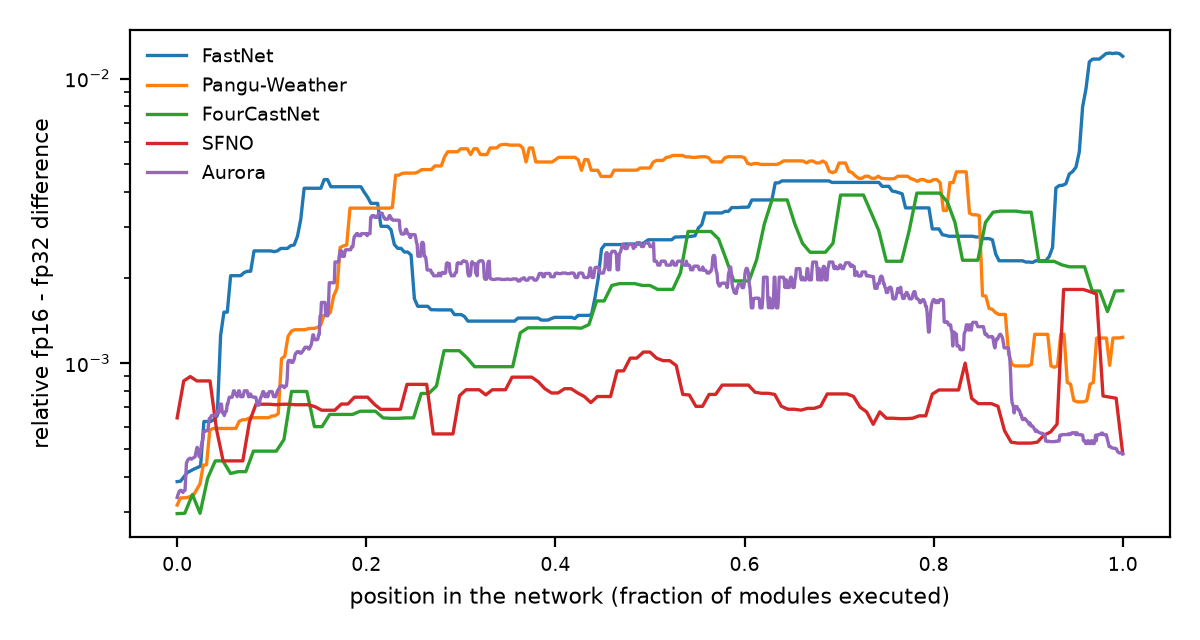}
\caption{Relative difference between the fp16 and fp32 outputs of every
module during the first model step, against the module's position in the
order of execution (median over 13 dates, running median over 1/40 of the
network).}
\label{fig:fp16-depth}
\end{figure}

Within a single step (Figure~\ref{fig:fp16-depth}), the
relative difference between fp16 and fp32 activations grows from about
$3\times10^{-4}$ after the first layers to a few $10^{-3}$, but not
monotonically: in Pangu-Weather and Aurora it falls again in the last stages, to about $10^{-3}$ and $5\times10^{-4}$, while in FastNet it rises steeply in the decoder, to about
$10^{-2}$ -- the component whose fp16 rounding dominates its departure.
FourCastNet's grows most steadily with depth. SFNO's stays between about
$5\times10^{-4}$ and $10^{-3}$ through the network, the smallest of the
five models, with a brief rise to $2\times10^{-3}$ near the end.

Large language models have a few activation channels with
values far larger than the rest, which complicate quantisation
\citep{dettmers2022llmint8}. The weather models show a similar but much
rarer pattern: the largest channel of a module exceeds the median channel
by more than a factor of 100 in 1 of 274 modules in Pangu-Weather, 3 of 98 in FourCastNet, 5 of 283 in FastNet and 6 of 1157 in Aurora (up to a factor of 660), and in none of SFNO's 132 (at most a factor of 63). These
values stay within fp16's numeric range (\S\ref{sec:results-headroom}). Nor do
they mark the sensitive parts of the network: they lie in sensitive
components in some cases -- the first stage of Pangu-Weather, FastNet's
decoder, FourCastNet's first blocks -- but in insensitive ones in others,
such as FastNet's mesh edge embedder (4\% of the departure when only it
runs in fp16) and
the second backbone stage of Aurora (6--14\%).

\subsection{Alternative quantisations: bf16 and 8-bit formats}
\label{sec:results-bf16}
\label{sec:results-alternatives}

We compare fp16 with two alternatives: the other 16-bit format, bf16, and
quantisation of the network to 8 bits.

bf16 is often preferred to fp16 because it keeps fp32's range.
With the state in fp32, that range is not needed: no fp16 value
overflowed, with a margin of at least a factor of 38
(\S\ref{sec:results-headroom}). bf16 runs as fast as fp16 and uses as
much memory and energy -- its time per step, peak memory and energy
differ from fp16's by at most 2\%, 1\% and 3.4\%
(Table~\ref{tab:benchmark}) -- but it is less
accurate (Table~\ref{tab:bf16}, Appendix~\ref{sec:appendix-bf16}). On
the same 52 dates and with the same fp32 state, bf16 raises the global
RMSE significantly more than fp16 in 10 to 20 of the 20 combinations of
variable and lead time per model (days 1, 3, 5 and 10), and is significantly
better in 2 of all 100, about as many as expected by chance. At days 1--5, the bf16
penalty in z500, 2t and msl is up to 3\% for Pangu-Weather, FastNet and SFNO, 5--25\% for
Aurora and 8--66\% for FourCastNet, whose bf16 2t error at day 10 is 1.8
times that of fp32 on the median date. This is what the formats predict:
bf16 has three fewer mantissa bits than fp16, so that its rounding steps
are eight times as large, and SFNO's bf16 forecasts depart from fp32 8 to
17 times as much as its fp16 forecasts at days 1--5. For these models,
fp16 is the better 16-bit format.

\label{sec:results-8bit}

\begin{table}[tbp]
\centering
\scriptsize
\setlength{\tabcolsep}{3pt}
\caption{Change in global RMSE (\%) relative to fp32 of the fp16 network and of 8-bit quantisation simulated inside it (weights only, or weights and the inputs of every linear and convolution layer; \S\ref{sec:methods-further}), on the 13 sensitivity dates. Bold: significant at the 5\% level (paired $t$-test).}
\label{tab:eightbit}
\begin{tabular}{llrrrrrr}
\toprule
 & & \multicolumn{3}{c}{z500} & \multicolumn{3}{c}{2t} \\
Model & Configuration & d1 & d5 & d10 & d1 & d5 & d10 \\
\midrule
FastNet & fp16 & \textbf{-0.4} & +0.4 & +0.2 & \textbf{-0.1} & -0.1 & +0.1 \\
 & int8 weights & \textbf{+171} & \textbf{+127} & \textbf{+72.4} & \textbf{+106} & \textbf{+380} & \textbf{+414} \\
 & fp8 weights & \textbf{+82.4} & \textbf{+102} & \textbf{+46.3} & \textbf{+17.9} & \textbf{+101} & \textbf{+91.5} \\
 & int8 weights + inputs & \textbf{+275} & \textbf{+144} & \textbf{+69.9} & \textbf{+142} & \textbf{+460} & \textbf{+409} \\
 & fp8 weights + inputs & \textbf{+155} & \textbf{+105} & \textbf{+40.4} & \textbf{+36.7} & \textbf{+121} & \textbf{+103} \\
\midrule
Pangu-Weather & fp16 & \textbf{+0.0} & +0.1 & \textbf{+0.4} & \textbf{+0.0} & \textbf{+0.1} & \textbf{+0.3} \\
 & int8 weights & \textbf{+5.5} & \textbf{+2.4} & +0.0 & \textbf{+1.4} & \textbf{+3.4} & +1.9 \\
 & fp8 weights & \textbf{+34.4} & \textbf{+15.9} & \textbf{+10.4} & \textbf{+13.7} & \textbf{+25.2} & \textbf{+24.0} \\
 & int8 weights + inputs & \textbf{+293} & \textbf{+90.9} & \textbf{+26.1} & \textbf{+101} & \textbf{+115} & \textbf{+77.3} \\
 & fp8 weights + inputs & \textbf{+148} & \textbf{+60.5} & \textbf{+21.4} & \textbf{+38.3} & \textbf{+39.3} & \textbf{+26.3} \\
\midrule
FourCastNet & fp16 & \textbf{+1.2} & \textbf{+0.5} & \textbf{-1.1} & \textbf{+0.2} & \textbf{+0.8} & +4.3 \\
 & int8 weights & \textbf{+15.6} & \textbf{+7.3} & +1.4 & \textbf{+3.4} & \textbf{+6.8} & \textbf{+55.4} \\
 & fp8 weights & \textbf{+65.6} & \textbf{+35.7} & \textbf{+18.4} & \textbf{+16.5} & \textbf{+49.2} & \textbf{+245} \\
 & int8 weights + inputs & \textbf{+33.0} & \textbf{+13.4} & \textbf{+1784} & \textbf{+11.0} & \textbf{+11.3} & \textbf{+12534} \\
 & fp8 weights + inputs & \textbf{+146} & \textbf{+54.0} & \textbf{+128} & \textbf{+43.1} & \textbf{+63.6} & \textbf{+1403} \\
\midrule
SFNO & fp16 & \textbf{+0.0} & \textbf{+0.1} & +0.1 & \textbf{+0.0} & \textbf{+0.1} & +0.1 \\
 & int8 weights & \textbf{+7.5} & \textbf{+15.4} & \textbf{+8.7} & \textbf{+8.8} & \textbf{+14.1} & \textbf{+7.1} \\
 & fp8 weights & \textbf{+20.1} & \textbf{+6.0} & \textbf{+5.1} & \textbf{+17.1} & \textbf{+17.4} & \textbf{+12.8} \\
 & int8 weights + inputs & \textbf{+432} & \textbf{+112} & \textbf{+34.4} & \textbf{+241} & \textbf{+147} & \textbf{+79.7} \\
 & fp8 weights + inputs & \textbf{+63.0} & \textbf{+10.3} & \textbf{+4.8} & \textbf{+31.9} & \textbf{+18.0} & \textbf{+7.3} \\
\midrule
Aurora & fp16 & \textbf{+0.2} & +0.1 & +0.1 & \textbf{+0.1} & +0.2 & +0.3 \\
 & int8 weights & \textbf{+1.7} & \textbf{+2.1} & +3.7 & \textbf{+1.1} & \textbf{+2.5} & \textbf{+3.1} \\
 & fp8 weights & \textbf{+40.8} & \textbf{+7.2} & -0.2 & \textbf{+7.4} & \textbf{+10.3} & \textbf{+7.5} \\
 & int8 weights + inputs & \textbf{+19.0} & \textbf{+9.1} & +1.2 & \textbf{+7.9} & \textbf{+5.8} & +3.8 \\
 & fp8 weights + inputs & \textbf{+179} & \textbf{+34.7} & +1.0 & \textbf{+33.1} & \textbf{+29.8} & \textbf{+10.0} \\
\bottomrule
\end{tabular}
\end{table}

Halving the precision once more, to 8 bits, is not a free
extension of fp16 (Table~\ref{tab:eightbit}). With int8 weights, the
mildest configuration, the global RMSE of z500 and 2t increases by
1--16\% at day 1 for Aurora, Pangu-Weather, FourCastNet and SFNO -- ten
to a hundred times the change in fp16 -- and mostly remains higher at
day 10 (up to 55\%, FourCastNet's 2t). fp8 weights, with only three
mantissa bits, cost more (7--82\% at day 1). Rounding the inputs of every
linear layer as well raises the day-1 RMSE by 8--430\%, and makes
FourCastNet's int8 forecasts unstable by day 10 (z500 RMSE up by a factor
of 19). By day 10, the 8-bit forecasts depart from the fp32 forecasts by
half the fp32 error or more, in most configurations by about the whole
error, against 3--40\% for fp16.

FastNet is the exception among the weight-only results: int8
weights more than double its z500 error, more than fp8 weights do.
Applying int8 to one layer at a time locates almost the whole effect in
the first layer of its encoder, which maps the 86 features of each
grid-to-mesh edge to the network's width: one of the 86 inputs has
weights about 75 times larger than a typical input, so that with one
scale per output channel the int8 steps are set by it in most channels, and about a third
of the layer's weights -- for ten of the inputs, most of them -- are
rounded to zero. fp8, whose steps are relative to each value, keeps them.
This is the kind of outlier that per-input scaling and the methods
developed for language models address
\citep{dettmers2022llmint8, xiao2022smoothquant}, and such methods, or
quantisation-aware fine-tuning, may make 8-bit inference usable for these
models. Our results show only that simple 8-bit quantisation is not a
drop-in replacement for fp32 as fp16 is; together with the cost of 8-bit
inference in our tests with Aurora (slower than fp16 and bf16, or out of
memory), they leave fp16 as the
practical choice for these models on current hardware.

\section{Discussion}
\label{sec:discussion}

For running AI weather models on a modest research setup with a single consumer-grade GPU, our results
support running the network in fp16 with the physical state,
normalisation and denormalisation in fp32. The change in skill is
statistically detectable with 52 paired dates but small in practical
terms -- less than an hour of lead time up to day 8 -- whereas the
savings in time and energy are a factor of two to four for four of
the five models, and about 1.4 for SFNO. Two conditions
come with this recommendation. First, the conversion must leave the
state in fp32: the one-line conversion of a whole model does not, and
for some models it fails silently, as for SFNO, whose complex weights
lose their imaginary parts. Second, fp16's numeric range
must be checked for each model, as we did with hooks on every module
(\S\ref{sec:results-headroom}); the margins we found differ by more than
a factor of ten between models, and a model trained or fine-tuned
differently may have larger activations.

Across the five models, which span 62 million to 1.26
billion parameters (Table~\ref{tab:models}), the sensitivity to fp16
follows neither model size nor fp32 skill. The two smallest models,
FastNet and FourCastNet, depart furthest from their fp32 forecasts
(Figure~\ref{fig:fp16-departure}), but Pangu-Weather and SFNO, of
intermediate size, depart least, and Aurora, the largest, lies in
between. SFNO has among the largest z500 errors of the five models yet is the
least affected by fp16, while Pangu-Weather and Aurora, the most skilful,
are affected little and moderately. The one model clearly more affected by fp16, FourCastNet, does not stand
out by size or skill: it has the largest changes in RMSE (up to 1.4\%,
against about 0.5\% at most for the others), the largest departure from
its fp32 forecasts relative to its error at day 1 (8\% in the mean over variables, against
1--5\%) and a tropical z500 penalty of up to 12\%
(\S\ref{sec:results-fp16}, \S\ref{sec:results-where}). A plausible cause is its
architecture, in particular its adaptive Fourier filters on the flat
latitude--longitude grid, but we have not isolated it; SFNO, which also
works in spectral space but on the sphere, is the least sensitive of the
five. With five models, this is an observation rather than a
test: the models also differ in step length (Pangu-Weather rounds 10
times in a 10-day forecast, the others 40 times), in which operations
always stay in fp32 (SFNO's spherical harmonic transforms), in
resolution, and in how fast they amplify small perturbations, all of
which affect how far the fp16 forecasts drift.

The computing power used for machine learning is increasingly
concentrated in industry, and academic groups have become less
represented in compute-intensive research
\citep{ahmed2020dedemocratization, besiroglu2024compute}. Hardware with
more memory has also become harder to afford: retail prices of consumer GPUs
have risen sharply, and most for the models with the most memory
\citep{techtimes2026gpu}, as graphics memory is increasingly taken up by AI
hardware \citep{trendforce2026memory}. This study also shows what direct access to
a modest local GPU offers over rented or remote hardware. Short tests
between experiments, such as those that located a memory leak in SFNO's
software or FastNet's sensitive layer in int8, took minutes and could
follow one another without waiting in a job queue; exclusive use of the GPU made timings reproducible
and allowed the board power to be sampled every 50\,ms, which shared or
virtualised GPUs often do not permit; the input data and the software
environments of five models stayed in place between experiments; and
the GPU cost nothing while it was idle. Together, these considerations
make it worthwhile to make AI weather models, efficient as they already
are, more efficient still: in fp16, every model studied here runs on a
single 16\,GB consumer GPU, Aurora without streaming its weights from CPU
memory, 1.3 to 3.5 times faster than in fp32 and at a fraction of the
energy, with practically no loss of skill.

The results are subject to several limitations. They cover five
models, chosen because they can be run in PyTorch on a 16\,GB GPU; larger
models, and models such as GraphCast that run in other frameworks, may
behave differently. All forecasts start in 2024, a single year, and are
verified against ERA5 rather than observations, for a few variables at
daily lead times up to 10 days. The storms are four cases and the
ensemble one model on 13 dates; they show that fp16 does not visibly
degrade these forecasts, not that it never does. The costs were measured
on one consumer GPU with one version of PyTorch, and both the speed-up
and the size of the rounding errors depend on the hardware and on the
kernels the library selects. Finally, the sensitivity experiments
measure departures from the fp32 forecast over the first day, on 13
dates; they rank the parts of each network but do not show how errors
introduced there affect skill at longer lead times.

The recommendation assumes hardware with fast fp16 arithmetic
(\S\ref{sec:quantisation}). Google's TPUs, on which several AI
weather models are developed, multiply matrices in bf16
\citep{wang2019bfloat16}; there bf16 is the natural 16-bit format, and
our results suggest it costs more skill (\S\ref{sec:results-bf16}). The
sensitivity experiments indicate where to look for a remedy: for
Pangu-Weather, Aurora and FastNet, the first or last components cause
most of the departure (\S\ref{sec:results-sensitivity}), so keeping these
in fp32 should remove much of it; for FourCastNet, whose sensitivity is
spread over the whole network, it may not.

Simple 8-bit quantisation, with one scale per output channel,
costs far more skill than fp16 (\S\ref{sec:results-8bit}). Newer formats
share a scale among small blocks of values \citep{rouhani2023microscaling},
and current GPUs, including the consumer GPU used here, compute in
floating-point formats down to 4 bits \citep{nvidia2025rtx50}. Block
scaling addresses the failure we found in FastNet, where a single input
with large weights set the int8 steps for most rows of a layer
(\S\ref{sec:results-8bit}), because each value then shares its scale only
with a few neighbours. Whether such formats, perhaps with quantisation-aware
fine-tuning, make 8- or 4-bit inference of weather models as accurate as
fp16 is open, as is whether they would be faster for models such as
Aurora, whose memory and time are dominated by intermediate results
rather than weights. The paired
design and module hooks used here would answer the first question
directly.

\section{Conclusion}
\label{sec:conclusion}

Running the network of an AI weather model in fp16, with its
physical state, normalisation and denormalisation kept in fp32, changes
its forecasts by much less than they differ from the truth: for five
models from three architecture families, the global RMSE changes by at
most 1.4\%, equivalent to at most two hours of lead time at day 10, and
the fp16 forecasts are neither smoother nor less calibrated, nor worse
for the storms examined. At the same time, a forecast is 1.3 to 3.5
times faster and uses 1.4 to 3.7 times less energy.
fp16's limited numeric range, its main risk, left a margin of at least a factor
of 38 in every model; this margin is model-specific and easy to check
with the hooks used here, and should be checked for any model before
fp16 is used. Where in the network the rounding matters most depends on
the architecture rather than on general rules, but since the effect of
the whole fp16 network is already small, no part of the network needs
to be protected. We therefore recommend fp16 inference, with the state
in fp32, as the default for research use of AI weather models on a
single GPU. bf16, although it needs no attention to range, offers no
advantage in speed and loses more skill, and simple 8-bit quantisation
is not a substitute.

\section*{Data and Code Availability}
All code is available at \url{https://github.com/sieste/ai-nwp-fp16}
(archived at Zenodo, \citealp{siegert2026code}). Its Python package \texttt{nwpprec} runs
the forecasts and experiments and produces the figures and tables in this
paper, apart from the tests of 8-bit inference with \texttt{torchao} on Aurora
and the layer-by-layer int8 test of FastNet; the
repository also holds the exact software environments, a check that the
stored results are reproduced, and the source of this paper. The scores
and other results behind the figures and tables are archived at Zenodo
\citep{siegert2026data}. ERA5 is available from the Copernicus Climate Data
Store, and the model checkpoints from their releases, listed below.

The author used a generative AI model (Anthropic Claude Sonnet 5.5 and Opus 5.5) to write and test
code, run experiments, and draft text of this paper. The author designed the research, directed and
checked the work, reviewed and revised all text, and takes full responsibility for the content.

The checkpoints are used under the terms stated by their
releases (accessed October 2026; a summary, not a legal reading).
Aurora~1.5 (deterministic and ensemble): MIT licence, for the weights on
Hugging Face (\texttt{microsoft/aurora}) and the \texttt{microsoft-aurora}
code. Pangu-Weather: the official weights are released under CC~BY-NC-SA~4.0,
and their commercial use is forbidden; the PyTorch reimplementation's
re-hosted weights (\texttt{zhaoshan/pangu\_pytorch}) are tagged
Apache-2.0 but are derived from the official ones, so we treat them as
non-commercial too, and the reimplementation's code carries no licence and
is not redistributed in our repository. FourCastNet: weights
(\texttt{nvidia/fourcastnet1} on Hugging Face) under Apache-2.0, loaded
with the Apache-2.0-licensed \texttt{earth2studio}. SFNO: checkpoint
(\texttt{sfno\_73ch\_small} on NVIDIA NGC) under Apache-2.0, run with
\texttt{earth2studio}, \texttt{makani} (Apache-2.0) and
\texttt{torch-harmonics} (BSD-3-Clause). FastNet: weights British Crown
copyright 2025, the Met Office, under the Open Government Licence~v3.0
(the Hugging Face card's metadata tag says CC~BY~4.0); its inference code,
which our FastNet adapter imports, is under the GNU Affero General Public
License~v3.0, so that adapter (\texttt{nwpprec/models/fastnet/}) is
licensed under it too; the rest of our code is under the MIT licence. AIFS, mentioned in \S\ref{sec:models}: weights
under CC~BY~4.0. All of these permit research use and the publication of
results derived from the models; the Pangu-Weather weights restrict use to
non-commercial purposes. ERA5 is distributed by the Copernicus Climate
Change Service under CC~BY~4.0 \citep{hersbach2020era5}. Contains modified Copernicus Climate
Change Service data (2024), under CC~BY~4.0. Neither the European Commission
nor ECMWF is responsible for any use that may be made of the data it contains.

\section*{Acknowledgments}
I thank Harris Sop Nkuiate for useful discussions throughout this project.

\bibliographystyle{plainnat}
\bibliography{references}

\begin{thebibliography}{70}
\providecommand{\natexlab}[1]{#1}
\providecommand{\url}[1]{\texttt{#1}}
\expandafter\ifx\csname urlstyle\endcsname\relax
  \providecommand{\doi}[1]{doi: #1}\else
  \providecommand{\doi}{doi: \begingroup \urlstyle{rm}\Url}\fi

\bibitem[Ahmed and Wahed(2020)]{ahmed2020dedemocratization}
Nur Ahmed and Muntasir Wahed.
\newblock The de-democratization of {AI}: Deep learning and the compute divide
  in artificial intelligence research.
\newblock arXiv preprint
  \href{https://arxiv.org/abs/2010.15581}{arXiv:2010.15581}, 2020.

\bibitem[Bauer et~al.(2015)Bauer, Thorpe, and Brunet]{bauer2015quiet}
Peter Bauer, Alan Thorpe, and Gilbert Brunet.
\newblock The quiet revolution of numerical weather prediction.
\newblock \emph{Nature}, 525\penalty0 (7567):\penalty0 47--55, 2015.
\newblock \doi{10.1038/nature14956}.

\bibitem[Ben~Bouall{\`e}gue et~al.(2024)Ben~Bouall{\`e}gue, Clare, Magnusson,
  et~al.]{benbouallegue2024rise}
Zied Ben~Bouall{\`e}gue, Mariana C.~A. Clare, Linus Magnusson, et~al.
\newblock The rise of data-driven weather forecasting: A first statistical
  assessment of machine learning--based weather forecasts in an
  operational-like context.
\newblock \emph{Bulletin of the American Meteorological Society}, 105\penalty0
  (6):\penalty0 E864--E883, 2024.
\newblock \doi{10.1175/BAMS-D-23-0162.1}.

\bibitem[Besiroglu et~al.(2024)Besiroglu, Bergerson, Michael,
  et~al.]{besiroglu2024compute}
Tamay Besiroglu, Sage~Andrus Bergerson, Amelia Michael, et~al.
\newblock The compute divide in machine learning: A threat to academic
  contribution and scrutiny?
\newblock arXiv preprint
  \href{https://arxiv.org/abs/2401.02452}{arXiv:2401.02452}, 2024.

\bibitem[Beven et~al.(2025{\natexlab{a}})Beven, Alaka, and
  Fritz]{beven2025milton}
John~L. Beven, II, Laura Alaka, and Cody Fritz.
\newblock Hurricane {Milton} ({AL142024}), 5--10 october 2024.
\newblock Tropical cyclone report, National Hurricane Center,
  2025{\natexlab{a}}.
\newblock \url{https://www.nhc.noaa.gov/data/tcr/AL142024_Milton.pdf}.

\bibitem[Beven et~al.(2025{\natexlab{b}})Beven, Fritz, and
  Alaka]{beven2025beryl}
John~L. Beven, II, Cody Fritz, and Laura Alaka.
\newblock Hurricane {Beryl} ({AL022024}), 28 june--9 july 2024.
\newblock Tropical cyclone report, National Hurricane Center,
  2025{\natexlab{b}}.
\newblock \url{https://www.nhc.noaa.gov/data/tcr/AL022024_Beryl.pdf}.

\bibitem[Bhattacharya et~al.(2026)Bhattacharya, Bhattacharya, and
  Jablonowski]{bhattacharya2026ptq}
Ananyo Bhattacharya, Swastik Bhattacharya, and Christiane Jablonowski.
\newblock Post-training quantization of autoregressive weather models.
\newblock \emph{arXiv preprint
  \href{https://arxiv.org/abs/2610.02511}{arXiv:2610.02511}}, 2026.

\bibitem[Bi et~al.(2023)Bi, Xie, Zhang, et~al.]{bi2023pangu}
Kaifeng Bi, Lingxi Xie, Hengheng Zhang, et~al.
\newblock Accurate medium-range global weather forecasting with 3d neural
  networks.
\newblock \emph{Nature}, 619:\penalty0 533--538, 2023.
\newblock \doi{10.1038/s41586-023-06185-3}.

\bibitem[Bloom(1995)]{bloom1995minimum}
Howard~S. Bloom.
\newblock Minimum detectable effects: A simple way to report the statistical
  power of experimental designs.
\newblock \emph{Evaluation Review}, 19\penalty0 (5):\penalty0 547--556, 1995.
\newblock \doi{10.1177/0193841X9501900504}.

\bibitem[Bodnar et~al.(2025)Bodnar, Bruinsma, Lucic, et~al.]{bodnar2024aurora}
Cristian Bodnar, Wessel~P. Bruinsma, Ana Lucic, et~al.
\newblock A foundation model for the {Earth} system.
\newblock \emph{Nature}, 641\penalty0 (8065):\penalty0 1180--1187, 2025.
\newblock \doi{10.1038/s41586-025-09005-y}.
\newblock Preprint: \href{https://arxiv.org/abs/2405.13063}{arXiv:2405.13063}.

\bibitem[Bonev et~al.(2023)Bonev, Kurth, Hundt, et~al.]{bonev2023sfno}
Boris Bonev, Thorsten Kurth, Christian Hundt, et~al.
\newblock Spherical {Fourier} neural operators: Learning stable dynamics on the
  sphere.
\newblock In \emph{Proceedings of the 40th International Conference on Machine
  Learning (ICML)}, 2023.
\newblock URL \url{https://proceedings.mlr.press/v202/bonev23a.html}.

\bibitem[Chantry et~al.(2019)Chantry, Thornes, Palmer, and
  D{\"u}ben]{chantry2019scale}
Matthew Chantry, Tobias Thornes, Tim Palmer, and Peter D{\"u}ben.
\newblock Scale-selective precision for weather and climate forecasting.
\newblock \emph{Monthly Weather Review}, 147\penalty0 (2):\penalty0 645--655,
  2019.
\newblock \doi{10.1175/MWR-D-18-0308.1}.

\bibitem[Choi et~al.(2018)Choi, Wang, Venkataramani, et~al.]{choi2018pact}
Jungwook Choi, Zhuo Wang, Swagath Venkataramani, et~al.
\newblock {PACT}: Parameterized clipping activation for quantized neural
  networks.
\newblock \emph{arXiv preprint
  \href{https://arxiv.org/abs/1805.06085}{arXiv:1805.06085}}, 2018.

\bibitem[Cohen(1988)]{cohen1988statistical}
Jacob Cohen.
\newblock \emph{Statistical Power Analysis for the Behavioral Sciences}.
\newblock Lawrence Erlbaum Associates, Hillsdale, NJ, 2nd edition, 1988.

\bibitem[Daub et~al.(2025)Daub, Dunstan, Bennett, et~al.]{daub2025fastnet}
Eric~G. Daub, Tom Dunstan, Thusal Bennett, et~al.
\newblock Technical overview and architecture of the {FastNet} machine learning
  weather prediction model, version 1.0.
\newblock arXiv preprint
  \href{https://arxiv.org/abs/2509.17658}{arXiv:2509.17658}, 2025.

\bibitem[Dettmers et~al.(2022)Dettmers, Lewis, Belkada, and
  Zettlemoyer]{dettmers2022llmint8}
Tim Dettmers, Mike Lewis, Younes Belkada, and Luke Zettlemoyer.
\newblock {LLM.int8()}: 8-bit matrix multiplication for transformers at scale.
\newblock \emph{arXiv preprint
  \href{https://arxiv.org/abs/2208.07339}{arXiv:2208.07339}}, 2022.

\bibitem[D{\"u}ben and Palmer(2014)]{dueben2014benchmark}
Peter~D. D{\"u}ben and T.~N. Palmer.
\newblock Benchmark tests for numerical weather forecasts on inexact hardware.
\newblock \emph{Monthly Weather Review}, 142\penalty0 (10):\penalty0
  3809--3829, 2014.
\newblock \doi{10.1175/MWR-D-14-00110.1}.

\bibitem[Dunstan et~al.(2025)Dunstan, Strickson, Bennett,
  et~al.]{dunstan2025fastnet}
Tom Dunstan, Oliver Strickson, Thusal Bennett, et~al.
\newblock {FastNet}: Improving the physical consistency of machine-learning
  weather prediction models through loss function design.
\newblock arXiv preprint
  \href{https://arxiv.org/abs/2509.17601}{arXiv:2509.17601}, 2025.

\bibitem[{ECMWF}(2026{\natexlab{a}})]{anemoi_inference_precision}
{ECMWF}.
\newblock Anemoi inference documentation: Optimisation and performance.
\newblock
  \url{https://anemoi.readthedocs.io/projects/inference/en/latest/usage/optimisation.html},
  2026{\natexlab{a}}.
\newblock Accessed 2026-10-08.

\bibitem[{ECMWF}(2026{\natexlab{b}})]{ecmwf2026aifscard}
{ECMWF}.
\newblock {AIFS Single v2} model card.
\newblock Hugging Face, \url{https://huggingface.co/ecmwf/aifs-single-2.0},
  2026{\natexlab{b}}.

\bibitem[Ferro(2014)]{ferro2014fair}
C.~A.~T. Ferro.
\newblock Fair scores for ensemble forecasts.
\newblock \emph{Quarterly Journal of the Royal Meteorological Society},
  140\penalty0 (683):\penalty0 1917--1923, 2014.
\newblock \doi{10.1002/qj.2270}.

\bibitem[Fuhrer et~al.(2018)Fuhrer, Chadha, Hoefler, et~al.]{fuhrer2018near}
Oliver Fuhrer, Tarun Chadha, Torsten Hoefler, et~al.
\newblock Near-global climate simulation at 1\,km resolution: establishing a
  performance baseline on 4888 {GPUs} with {COSMO} 5.0.
\newblock \emph{Geoscientific Model Development}, 11\penalty0 (4):\penalty0
  1665--1681, 2018.
\newblock \doi{10.5194/gmd-11-1665-2018}.

\bibitem[Giorgetta et~al.(2022)Giorgetta, Sawyer, Lapillonne,
  et~al.]{giorgetta2022icon}
Marco~A. Giorgetta, William Sawyer, Xavier Lapillonne, et~al.
\newblock The {ICON-A} model for direct {QBO} simulations on {GPUs} (version
  icon-cscs:baf28a514).
\newblock \emph{Geoscientific Model Development}, 15\penalty0 (18):\penalty0
  6985--7016, 2022.
\newblock \doi{10.5194/gmd-15-6985-2022}.

\bibitem[{Google DeepMind}(2023)]{deepmind_weathernext_casting}
{Google DeepMind}.
\newblock {WeatherNext} / {GraphCast} source code:
  \texttt{weathernext/utils/casting.py} ({BFloat16} casting wrapper).
\newblock \url{https://github.com/google-deepmind/weathernext}, 2023.
\newblock Accessed 2026-10-08.

\bibitem[Guibas et~al.(2022)Guibas, Mardani, Li, et~al.]{guibas2022afno}
John Guibas, Morteza Mardani, Zongyi Li, et~al.
\newblock Efficient token mixing for transformers via adaptive {Fourier} neural
  operators.
\newblock In \emph{International Conference on Learning Representations
  (ICLR)}, 2022.
\newblock URL \url{https://openreview.net/forum?id=EXHG-A3jlM}.

\bibitem[Hatfield et~al.(2018)Hatfield, Subramanian, Palmer, and
  D{\"u}ben]{hatfield2018improving}
Sam Hatfield, Aneesh Subramanian, Tim Palmer, and Peter D{\"u}ben.
\newblock Improving weather forecast skill through reduced-precision data
  assimilation.
\newblock \emph{Monthly Weather Review}, 146\penalty0 (1):\penalty0 49--62,
  2018.
\newblock \doi{10.1175/MWR-D-17-0132.1}.

\bibitem[Hatfield et~al.(2019)Hatfield, Chantry, D{\"u}ben, and
  Palmer]{hatfield2019accelerating}
Sam Hatfield, Matthew Chantry, Peter D{\"u}ben, and Tim Palmer.
\newblock Accelerating high-resolution weather models with deep-learning
  hardware.
\newblock In \emph{Proceedings of the Platform for Advanced Scientific
  Computing Conference (PASC '19)}, pages 1--11, 2019.
\newblock \doi{10.1145/3324989.3325711}.

\bibitem[Heimersheim and Nanda(2024)]{heimersheim2024activation}
Stefan Heimersheim and Neel Nanda.
\newblock How to use and interpret activation patching.
\newblock arXiv preprint
  \href{https://arxiv.org/abs/2404.15255}{arXiv:2404.15255}, 2024.

\bibitem[Hersbach et~al.(2020)Hersbach, Bell, Berrisford,
  et~al.]{hersbach2020era5}
Hans Hersbach, Bill Bell, Paul Berrisford, et~al.
\newblock The {ERA5} global reanalysis.
\newblock \emph{Quarterly Journal of the Royal Meteorological Society},
  146\penalty0 (730):\penalty0 1999--2049, 2020.
\newblock \doi{10.1002/qj.3803}.

\bibitem[Jacob et~al.(2018)Jacob, Kligys, Chen, et~al.]{jacob2018quantization}
Benoit Jacob, Skirmantas Kligys, Bo~Chen, et~al.
\newblock Quantization and training of neural networks for efficient
  integer-arithmetic-only inference.
\newblock In \emph{Proceedings of the IEEE Conference on Computer Vision and
  Pattern Recognition (CVPR)}, pages 2704--2713, 2018.
\newblock \doi{10.1109/CVPR.2018.00286}.

\bibitem[Kalamkar et~al.(2019)Kalamkar, Mudigere, Mellempudi,
  et~al.]{kalamkar2019bfloat16}
Dhiraj Kalamkar, Dheevatsa Mudigere, Naveen Mellempudi, et~al.
\newblock A study of {BFLOAT16} for deep learning training.
\newblock \emph{arXiv preprint
  \href{https://arxiv.org/abs/1905.12322}{arXiv:1905.12322}}, 2019.

\bibitem[Kl{\"o}wer et~al.(2022)Kl{\"o}wer, Hatfield, Croci, D{\"u}ben, and
  Palmer]{klower2022fluid}
Milan Kl{\"o}wer, Samuel Hatfield, Matteo Croci, Peter~D. D{\"u}ben, and Tim
  Palmer.
\newblock Fluid simulations accelerated with 16 bits: Approaching 4x speedup on
  {A64FX} by squeezing {ShallowWaters.jl} into {Float16}.
\newblock \emph{Journal of Advances in Modeling Earth Systems}, 14\penalty0
  (2):\penalty0 e2021MS002684, 2022.
\newblock \doi{10.1029/2021MS002684}.

\bibitem[Kochkov et~al.(2024)Kochkov, Yuval, Langmore,
  et~al.]{kochkov2024neural}
Dmitrii Kochkov, Janni Yuval, Ian Langmore, et~al.
\newblock Neural general circulation models for weather and climate.
\newblock \emph{Nature}, 632\penalty0 (8027):\penalty0 1060--1066, 2024.
\newblock \doi{10.1038/s41586-024-07744-y}.

\bibitem[Kurth et~al.(2023)Kurth, Subramanian, Harrington,
  et~al.]{kurth2023fourcastnet}
Thorsten Kurth, Shashank Subramanian, Peter Harrington, et~al.
\newblock {FourCastNet}: Accelerating global high-resolution weather
  forecasting using adaptive {Fourier} neural operators.
\newblock In \emph{Proceedings of the Platform for Advanced Scientific
  Computing Conference (PASC '23)}, pages 1--11, 2023.
\newblock \doi{10.1145/3592979.3593412}.

\bibitem[Lam et~al.(2023)Lam, Sanchez-Gonzalez, Willson,
  et~al.]{lam2023graphcast}
Remi Lam, Alvaro Sanchez-Gonzalez, Matthew Willson, et~al.
\newblock Learning skillful medium-range global weather forecasting.
\newblock \emph{Science}, 382\penalty0 (6677):\penalty0 1416--1421, 2023.
\newblock \doi{10.1126/science.adi2336}.

\bibitem[Lang et~al.(2024)Lang, Alexe, Chantry, et~al.]{lang2024aifs}
Simon Lang, Mihai Alexe, Matthew Chantry, et~al.
\newblock {AIFS} -- {ECMWF}'s data-driven forecasting system.
\newblock arXiv preprint
  \href{https://arxiv.org/abs/2406.01465}{arXiv:2406.01465}, 2024.

\bibitem[Lang et~al.(2021)Lang, Dawson, Diamantakis, et~al.]{lang2021more}
Simon T.~K. Lang, Andrew Dawson, Michail Diamantakis, et~al.
\newblock More accuracy with less precision.
\newblock \emph{Quarterly Journal of the Royal Meteorological Society},
  147\penalty0 (741):\penalty0 4358--4370, 2021.
\newblock \doi{10.1002/qj.4181}.

\bibitem[Lapillonne et~al.(2026)Lapillonne, Hupp, Gessler,
  et~al.]{lapillonne2026icon}
Xavier Lapillonne, Daniel Hupp, Fabian Gessler, et~al.
\newblock Operational numerical weather prediction with {ICON} on {GPUs}
  (version 2024.10).
\newblock \emph{Geoscientific Model Development}, 19\penalty0 (2):\penalty0
  755--772, 2026.
\newblock \doi{10.5194/gmd-19-755-2026}.

\bibitem[Leutbecher and Palmer(2008)]{leutbecher2008ensemble}
M.~Leutbecher and T.~N. Palmer.
\newblock Ensemble forecasting.
\newblock \emph{Journal of Computational Physics}, 227\penalty0 (7):\penalty0
  3515--3539, 2008.
\newblock \doi{10.1016/j.jcp.2007.02.014}.

\bibitem[Lin et~al.(2023)Lin, Tang, Tang, et~al.]{lin2023awq}
Ji~Lin, Jiaming Tang, Haotian Tang, et~al.
\newblock {AWQ}: Activation-aware weight quantization for {LLM} compression and
  acceleration.
\newblock \emph{arXiv preprint
  \href{https://arxiv.org/abs/2306.00978}{arXiv:2306.00978}}, 2023.

\bibitem[Luccioni et~al.(2024)Luccioni, Jernite, and
  Strubell]{luccioni2024power}
Sasha Luccioni, Yacine Jernite, and Emma Strubell.
\newblock Power hungry processing: Watts driving the cost of {AI} deployment?
\newblock In \emph{Proceedings of the 2024 ACM Conference on Fairness,
  Accountability, and Transparency (FAccT '24)}, pages 85--99, 2024.
\newblock \doi{10.1145/3630106.3658542}.

\bibitem[Meng et~al.(2022)Meng, Bau, Andonian, and Belinkov]{meng2022locating}
Kevin Meng, David Bau, Alex Andonian, and Yonatan Belinkov.
\newblock Locating and editing factual associations in {GPT}.
\newblock In \emph{Advances in Neural Information Processing Systems},
  volume~35, 2022.
\newblock Preprint: \href{https://arxiv.org/abs/2202.05262}{arXiv:2202.05262}.

\bibitem[{Met Office}(2024{\natexlab{a}})]{metoffice2024isha}
{Met Office}.
\newblock Storm {Isha} has been named.
\newblock Press release, 19 January 2024,
  \url{https://www.metoffice.gov.uk/about-us/news-and-media/media-centre/weather-and-climate-news/2024/storm-isha-has-been-named},
  2024{\natexlab{a}}.

\bibitem[{Met Office}(2024{\natexlab{b}})]{metoffice2024kathleen}
{Met Office}.
\newblock Storm {Kathleen} named and weather warnings issued.
\newblock Press release, 4 April 2024,
  \url{https://www.metoffice.gov.uk/about-us/news-and-media/media-centre/weather-and-climate-news/2024/storm-kathleen-named-and-weather-warnings-issued},
  2024{\natexlab{b}}.

\bibitem[{Met Office}(2024{\natexlab{c}})]{metoffice2024storms}
{Met Office}.
\newblock Looking back on a storm-laden season.
\newblock Met Office blog,
  \url{https://www.metoffice.gov.uk/blog/2024/looking-back-on-a-storm-laden-season},
  2024{\natexlab{c}}.

\bibitem[Micikevicius et~al.(2018)Micikevicius, Narang, Alben,
  et~al.]{micikevicius2018mixed}
Paulius Micikevicius, Sharan Narang, Jonah Alben, et~al.
\newblock Mixed precision training.
\newblock \emph{arXiv preprint
  \href{https://arxiv.org/abs/1710.03740}{arXiv:1710.03740}}, 2018.

\bibitem[Micikevicius et~al.(2022)Micikevicius, Stosic, Burgess,
  et~al.]{micikevicius2022fp8}
Paulius Micikevicius, Dusan Stosic, Neil Burgess, et~al.
\newblock {FP8} formats for deep learning.
\newblock arXiv preprint
  \href{https://arxiv.org/abs/2209.05433}{arXiv:2209.05433}, 2022.

\bibitem[{Microsoft Research AI for Science}(2026)]{aurora_docs}
{Microsoft Research AI for Science}.
\newblock Aurora: A foundation model for the earth system -- documentation of
  the {Aurora}~1.5 checkpoints.
\newblock \url{https://microsoft.github.io/aurora/models.html}, 2026.
\newblock Package \texttt{microsoft-aurora} 2.0.1; accessed October 2026.

\bibitem[Nagel et~al.(2021)Nagel, Fournarakis, Amjad, et~al.]{nagel2021white}
Markus Nagel, Marios Fournarakis, Rana~Ali Amjad, et~al.
\newblock A white paper on neural network quantization.
\newblock arXiv preprint
  \href{https://arxiv.org/abs/2106.08295}{arXiv:2106.08295}, 2021.

\bibitem[{NVIDIA}(2017)]{nvidia2017volta}
{NVIDIA}.
\newblock {NVIDIA} {Tesla} {V100} {GPU} architecture.
\newblock Whitepaper WP-08608-001,
  \url{https://images.nvidia.com/content/volta-architecture/pdf/volta-architecture-whitepaper.pdf},
  2017.

\bibitem[{NVIDIA}(2020)]{nvidia2020ampere}
{NVIDIA}.
\newblock {NVIDIA} {A100} tensor core {GPU} architecture.
\newblock Whitepaper,
  \url{https://images.nvidia.com/aem-dam/en-zz/Solutions/data-center/nvidia-ampere-architecture-whitepaper.pdf},
  2020.

\bibitem[{NVIDIA}(2025)]{nvidia2025rtx50}
{NVIDIA}.
\newblock {NVIDIA} {Blackwell} {GeForce} {RTX} 50 series opens new world of
  {AI} computer graphics.
\newblock Press release, 6 January 2025,
  \url{https://nvidianews.nvidia.com/news/nvidia-blackwell-geforce-rtx-50-series-opens-new-world-of-ai-computer-graphics},
  2025.

\bibitem[Pathak et~al.(2022)Pathak, Subramanian, Harrington,
  et~al.]{pathak2022fourcastnet}
Jaideep Pathak, Shashank Subramanian, Peter Harrington, et~al.
\newblock {FourCastNet}: A global data-driven high-resolution weather model
  using adaptive fourier neural operators.
\newblock \emph{arXiv preprint
  \href{https://arxiv.org/abs/2202.11214}{arXiv:2202.11214}}, 2022.

\bibitem[Patterson et~al.(2022)Patterson, Gonzalez, H{\"o}lzle,
  et~al.]{patterson2022carbon}
David Patterson, Joseph Gonzalez, Urs H{\"o}lzle, et~al.
\newblock The carbon footprint of machine learning training will plateau, then
  shrink.
\newblock \emph{Computer}, 55\penalty0 (7):\penalty0 18--28, 2022.
\newblock \doi{10.1109/MC.2022.3148714}.

\bibitem[Paxton et~al.(2022)Paxton, Chantry, Kl{\"o}wer, Saffin, and
  Palmer]{paxton2022climate}
E.~Adam Paxton, Matthew Chantry, Milan Kl{\"o}wer, Leo Saffin, and Tim Palmer.
\newblock Climate modeling in low precision: Effects of both deterministic and
  stochastic rounding.
\newblock \emph{Journal of Climate}, 35\penalty0 (4):\penalty0 1215--1229,
  2022.
\newblock \doi{10.1175/JCLI-D-21-0343.1}.

\bibitem[Rasp et~al.(2020)Rasp, Dueben, Scher, et~al.]{rasp2020weatherbench}
Stephan Rasp, Peter~D. Dueben, Sebastian Scher, et~al.
\newblock {WeatherBench}: A benchmark dataset for data-driven weather
  forecasting.
\newblock \emph{Journal of Advances in Modeling Earth Systems}, 12\penalty0
  (11):\penalty0 e2020MS002203, 2020.
\newblock \doi{10.1029/2020MS002203}.

\bibitem[Rasp et~al.(2024)Rasp, Hoyer, Merose, et~al.]{rasp2024weatherbench2}
Stephan Rasp, Stephan Hoyer, Alexander Merose, et~al.
\newblock {WeatherBench} 2: A benchmark for the next generation of data-driven
  global weather models.
\newblock \emph{Journal of Advances in Modeling Earth Systems}, 16\penalty0
  (6):\penalty0 e2023MS004019, 2024.
\newblock \doi{10.1029/2023MS004019}.

\bibitem[Rouhani et~al.(2023)Rouhani, Zhao, More,
  et~al.]{rouhani2023microscaling}
Bita~Darvish Rouhani, Ritchie Zhao, Ankit More, et~al.
\newblock Microscaling data formats for deep learning.
\newblock arXiv preprint
  \href{https://arxiv.org/abs/2310.10537}{arXiv:2310.10537}, 2023.

\bibitem[Saltelli and Annoni(2010)]{saltelli2010perfunctory}
Andrea Saltelli and Paola Annoni.
\newblock How to avoid a perfunctory sensitivity analysis.
\newblock \emph{Environmental Modelling \& Software}, 25\penalty0
  (12):\penalty0 1508--1517, 2010.
\newblock \doi{10.1016/j.envsoft.2010.04.012}.

\bibitem[Siegert(2026{\natexlab{a}})]{siegert2026code}
Stefan Siegert.
\newblock Half-precision inference of {AI} weather prediction models: code.
\newblock Zenodo, version 1.0.1, \url{https://doi.org/10.5281/zenodo.24010291},
  2026{\natexlab{a}}.

\bibitem[Siegert(2026{\natexlab{b}})]{siegert2026data}
Stefan Siegert.
\newblock Results: Half-precision inference of {AI} weather prediction models.
\newblock Zenodo, version 1.0.0, \url{https://doi.org/10.5281/zenodo.23887144},
  2026{\natexlab{b}}.

\bibitem[Siegert et~al.(2011)Siegert, Br{\"o}cker, and
  Kantz]{siegert2011outliers}
Stefan Siegert, Jochen Br{\"o}cker, and Holger Kantz.
\newblock Predicting outliers in ensemble forecasts.
\newblock \emph{Quarterly Journal of the Royal Meteorological Society},
  137\penalty0 (660):\penalty0 1887--1897, 2011.
\newblock \doi{10.1002/qj.868}.

\bibitem[{Tech Times}(2026)]{techtimes2026gpu}
{Tech Times}.
\newblock {GPU} prices jump again as {AMD} {Radeon} loses its last
  affordability buffer.
\newblock 12 October 2026, reporting retail price data from 3DCenter,
  \url{https://www.techtimes.com/articles/328936/20261012/gpu-prices-jump-again-amd-radeon-loses-its-last-affordability-buffer.htm},
  2026.

\bibitem[{TrendForce}(2026)]{trendforce2026memory}
{TrendForce}.
\newblock {AI} server demand sustains memory contract price increases in
  {4Q26}, while consumer-side pressure persists.
\newblock Press release, 30 September 2026,
  \url{https://www.trendforce.com/presscenter/news/20260930-13258.html}, 2026.

\bibitem[V{\'a}{\v{n}}a et~al.(2017)V{\'a}{\v{n}}a, D{\"u}ben, Lang,
  et~al.]{vana2017single}
Filip V{\'a}{\v{n}}a, Peter D{\"u}ben, Simon Lang, et~al.
\newblock Single precision in weather forecasting models: An evaluation with
  the {IFS}.
\newblock \emph{Monthly Weather Review}, 145\penalty0 (2):\penalty0 495--502,
  2017.
\newblock \doi{10.1175/MWR-D-16-0228.1}.

\bibitem[Wang and Kanwar(2019)]{wang2019bfloat16}
Shibo Wang and Pankaj Kanwar.
\newblock {BFloat16}: The secret to high performance on {Cloud TPUs}.
\newblock Google Cloud blog, 23 August 2019,
  \url{https://cloud.google.com/blog/products/ai-machine-learning/bfloat16-the-secret-to-high-performance-on-cloud-tpus},
  2019.

\bibitem[Weyn et~al.(2026)Weyn, Ni, Misra, et~al.]{weyn2026aurora15}
Jonathan~A. Weyn, Zekun Ni, Amit Misra, et~al.
\newblock Aurora 1.5: Fine-tuning a foundation model for medium-range ensemble
  weather prediction.
\newblock Preprint, Microsoft,
  \url{https://www.microsoft.com/en-us/research/wp-content/uploads/2026/07/Aurora_1_5_Paper.pdf},
  2026.

\bibitem[{World Meteorological Organization}(2010)]{wmo2010gdpfs}
{World Meteorological Organization}.
\newblock \emph{Manual on the Global Data-processing and Forecasting System,
  Volume {I}: Global Aspects}.
\newblock WMO-No. 485. World Meteorological Organization, Geneva, 2010.

\bibitem[Xiao et~al.(2022)Xiao, Lin, Seznec, et~al.]{xiao2022smoothquant}
Guangxuan Xiao, Ji~Lin, Mickael Seznec, et~al.
\newblock {SmoothQuant}: Accurate and efficient post-training quantization for
  large language models.
\newblock \emph{arXiv preprint
  \href{https://arxiv.org/abs/2211.10438}{arXiv:2211.10438}}, 2022.

\bibitem[Zhao(2024)]{pangu_pytorch}
Shan Zhao.
\newblock pangu-pytorch: a {PyTorch} implementation of {Pangu-Weather} at the
  24-hour horizon.
\newblock \url{https://github.com/zhaoshan2/pangu-pytorch}, 2024.
\newblock Commit 97db96b; converted weights from
  \url{https://huggingface.co/zhaoshan/pangu_pytorch}, revision 65cd779.

\end{thebibliography}

\appendix
\section{Supplementary figure}
\label{sec:appendix-figures}

\begin{figure}[htbp]
\centering
\includegraphics[width=\textwidth]{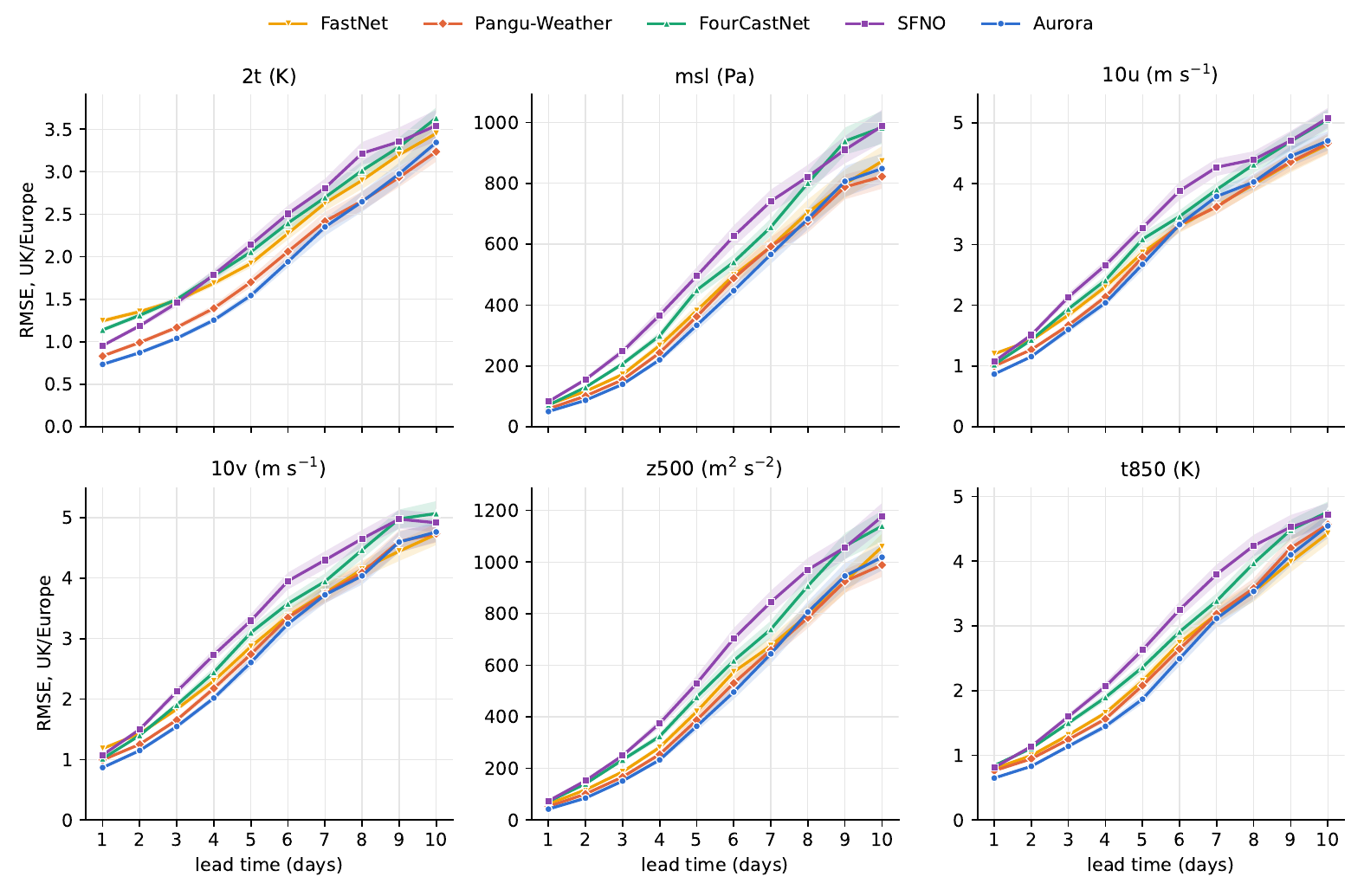}
\caption{As Figure~\ref{fig:model-skill}, for the UK/Europe region:
fp32 forecast skill of the models studied, mean RMSE against ERA5 vs.\
lead time.}
\label{fig:model-skill-uk}
\end{figure}

 Figure~\ref{fig:model-skill-uk} is the UK/Europe counterpart of Figure~\ref{fig:model-skill}.

\section{bf16 compared with fp16}
\label{sec:appendix-bf16}

\begin{figure}[tbp]
\centering
\includegraphics[width=\textwidth]{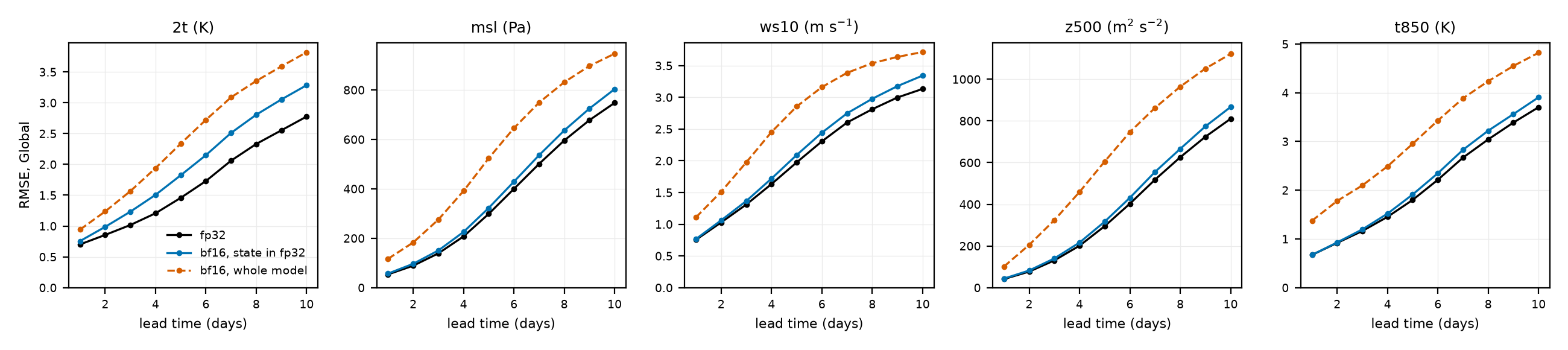}
\caption{Mean global RMSE against ERA5 of Aurora's fp32 forecasts, of its
forecasts with the network in bf16 and the state in fp32, and of its
forecasts with the whole model converted with
\texttt{model.to(torch.bfloat16)}, which also rounds the state in physical
units; 52 dates in 2024.}
\label{fig:rmse-grid-naive}
\end{figure}

Figure~\ref{fig:rmse-grid-naive} shows the RMSE of Aurora's bf16 forecasts
when the whole model is converted, so that it also holds its state in bf16
in physical units (\S\ref{sec:results-state}), compared with the bf16
network with the state in fp32. The whole-model conversion raises the
RMSE by 19--167\%, the bf16 network alone by at most 25\%.

\begin{table}[htbp]
\centering
\scriptsize
\setlength{\tabcolsep}{3pt}
\caption{Change in global RMSE (\%) relative to fp32 with the network in bf16 and in fp16 (state in fp32 in both), 52 dates in 2024. Bold: significant at the 5\% level (paired $t$-test).}
\label{tab:bf16}
\begin{tabular}{llrrrrrrrrrrrr}
\toprule
 & & \multicolumn{4}{c}{z500} & \multicolumn{4}{c}{2t} & \multicolumn{4}{c}{msl} \\
Model & Precision & d1 & d3 & d5 & d10 & d1 & d3 & d5 & d10 & d1 & d3 & d5 & d10 \\
\midrule
FastNet & bf16 & \textbf{+1.1} & \textbf{+2.1} & \textbf{+3.0} & -0.6 & \textbf{+0.2} & \textbf{+0.6} & \textbf{+0.8} & -0.2 & \textbf{+1.0} & \textbf{+1.7} & \textbf{+1.9} & \textbf{-2.0} \\
 & fp16 & \textbf{-0.3} & \textbf{-0.3} & +0.1 & 0.0 & \textbf{-0.1} & \textbf{-0.2} & \textbf{-0.2} & -0.1 & \textbf{-0.1} & \textbf{-0.2} & +0.1 & +0.3 \\
\midrule
Pangu-Weather & bf16 & \textbf{+3.1} & \textbf{+1.0} & \textbf{+0.6} & +0.7 & \textbf{+0.7} & \textbf{+0.3} & 0.0 & -0.6 & \textbf{+0.6} & \textbf{+0.5} & +0.4 & +0.4 \\
 & fp16 & \textbf{+0.1} & \textbf{+0.1} & \textbf{+0.1} & \textbf{+0.2} & \textbf{0.0} & \textbf{+0.1} & \textbf{+0.1} & \textbf{+0.2} & \textbf{0.0} & \textbf{+0.1} & \textbf{+0.1} & \textbf{+0.2} \\
\midrule
FourCastNet & bf16 & \textbf{+66.4} & \textbf{+28.9} & \textbf{+16.8} & \textbf{+28.5} & \textbf{+14.6} & \textbf{+19.4} & \textbf{+20.0} & \textbf{+271.4} & \textbf{+11.0} & \textbf{+10.2} & \textbf{+8.5} & \textbf{+23.0} \\
 & fp16 & \textbf{+1.4} & \textbf{+0.6} & \textbf{+0.2} & -0.2 & \textbf{+0.2} & \textbf{+0.4} & \textbf{+0.5} & -0.6 & 0.0 & \textbf{-0.2} & \textbf{-0.3} & \textbf{-0.7} \\
\midrule
SFNO & bf16 & \textbf{+0.9} & 0.0 & 0.0 & +0.1 & \textbf{+0.8} & \textbf{+1.0} & \textbf{+1.0} & +0.4 & \textbf{+2.0} & \textbf{+1.0} & \textbf{+0.4} & -0.1 \\
 & fp16 & \textbf{0.0} & \textbf{0.0} & \textbf{+0.1} & 0.0 & \textbf{0.0} & \textbf{0.0} & \textbf{0.0} & 0.0 & \textbf{0.0} & 0.0 & \textbf{0.0} & 0.0 \\
\midrule
Aurora & bf16 & \textbf{+5.4} & \textbf{+7.2} & \textbf{+7.3} & \textbf{+7.0} & \textbf{+7.7} & \textbf{+21.3} & \textbf{+25.2} & \textbf{+18.3} & \textbf{+6.3} & \textbf{+8.5} & \textbf{+8.1} & \textbf{+7.4} \\
 & fp16 & \textbf{+0.2} & \textbf{+0.1} & \textbf{+0.1} & \textbf{+0.3} & \textbf{+0.1} & \textbf{+0.1} & \textbf{+0.2} & +0.1 & \textbf{+0.1} & \textbf{+0.3} & \textbf{+0.2} & \textbf{+0.5} \\
\bottomrule
\end{tabular}
\end{table}

Table~\ref{tab:bf16} compares the change in global RMSE with the
network in bf16 and in fp16, the state in fp32 in both, on the 52 dates.
For every model, the bf16 penalty exceeds the fp16 change at almost
every lead time, by one to two orders of magnitude for Aurora and
FourCastNet. FourCastNet's bf16 forecasts degrade widely by day 10: its
2t error is more than twice that of fp32 on 21 of the 52 dates. Where
bf16 does better than fp32, the differences are small and, except for
FastNet's msl at day 10 ($-2.0\%$), not significant.

\section{Minimum detectable effect}
\label{sec:appendix-mde}

Let $d_i$ be the fp16 RMSE minus the fp32 RMSE on date $i$, with mean
$\bar d$ and standard deviation $s$ over the $N$ dates, and let
$t_{p,\nu}$ be the $p$-quantile of the $t$ distribution with $\nu$ degrees
of freedom. The two-sided paired $t$-test at the 5\% level finds a
significant change when $|\bar d|$ exceeds $t_{0.975,N-1}\,s/\sqrt{N}$.
If the true change is $\delta>0$, $\bar d$ scatters around $\delta$ with
standard error $s/\sqrt{N}$, and lies above
$\delta-t_{0.8,N-1}\,s/\sqrt{N}$ with probability 0.8. The test therefore
detects the change with probability 0.8 when this value equals the
significance threshold, which gives the minimum detectable effect
\begin{equation}
\delta=\left(t_{0.975,N-1}+t_{0.8,N-1}\right)\frac{s}{\sqrt{N}}.
\label{eq:mde}
\end{equation}
The first term is the significance threshold. A true change of that size
would be detected only about half the time, and the second term is the
margin that raises this to 80\%. The same holds for $\delta<0$ by symmetry.
For $N=52$, $t_{0.975,51}=2.008$ and $t_{0.8,51}=0.849$, so that
$\delta\approx0.40\,s$; Table~\ref{tab:power} gives $\delta$ as a percentage
of the fp32 RMSE.

Equation~\eqref{eq:mde} treats $\bar d$ as following a $t$ distribution
shifted by $\delta$, whereas the test statistic strictly follows a
noncentral $t$ distribution. For $N=52$ the exact calculation gives the
same value of $\delta$ to within 0.1\%. Like the test itself, the
calculation assumes that the $d_i$ are independent between dates, and it
treats the observed $s$ as the true standard deviation.

\end{document}